\documentclass[10pt]{iopart}

\usepackage{iopams}  
\usepackage{graphicx}
\usepackage{amssymb}
\usepackage{epstopdf}
\usepackage{float}
\usepackage{color}
\usepackage{subfigure}
\usepackage{relsize}
\expandafter\let\csname equation*\endcsname\relax

\expandafter\let\csname endequation*\endcsname\relax

\usepackage{amsmath}
\usepackage[ngerman, english]{babel}
\newcommand{\cev}[1]{\reflectbox{\ensuremath{\vec{\reflectbox{\ensuremath{#1}}}}}}

\begin{document}

\title{Ultracold molecular Rydberg physics in a high density environment}

\author{Matthew T Eiles$^1$, Jes\'{u}s P\'{e}rez-R\'{i}os$^1$, F Robicheaux$^{1,2}$, Chris H Greene$^{1,2}$}

\address{$^1$Department of Physics and Astronomy, Purdue University, West Lafayette, Indiana, 47907, USA.\newline
$^2$ Purdue Quantum Center, Purdue University, West Lafayette, Indiana, 47907, USA.}
\ead{\mailto{meiles@purdue.edu*},\mailto{jperezri@purdue.edu},\mailto{robichf@purdue.edu},\mailto{chgreene@purdue.edu}}

\vspace{10pt}
\begin{indented}
\item \today
\\
*corresponding author
\end{indented}

\begin{abstract}
Sufficiently high densities in Bose-Einstein condensates provide favorable conditions for the production of ultralong-range polyatomic molecules consisting of one Rydberg atom and a number of neutral ground state atoms. The chemical binding properties and electronic wave functions of these exotic molecules are investigated analytically via hybridized diatomic states. The effects of the molecular geometry on the system's properties are studied through comparisons of the adiabatic potential curves and electronic structures for both symmetric and randomly configured molecular geometries. General properties of these molecules with increasing numbers of constituent atoms and in different geometries are presented. These polyatomic states have spectral signatures that lead to non-Lorentzian line-profiles.
\end{abstract}

%

%
%
%
\ioptwocol

\newcommand{\be}{\begin{equation}}
\newcommand{\ee}{\end{equation}}
\newcommand{\lap}{\nabla^2}
\newcommand{\pd}[1]{\frac{\partial}{\partial{#1}}}
\newcommand{\pdd}[1]{\frac{\partial^2}{\partial{#1}^2}}
\newcommand{\pdde}[2]{\frac{\partial^2{#1}}{\partial{#2}^2}}
\newcommand{\pde}[2]{\frac{\partial{#1}}{\partial{#2}}}
\newcommand{\ar}{(\vec{r},t)}
\newcommand{\arz}{(\vec{r},0)}
\newcommand{\psirt}{\psi(\vec{r},t)}
\newcommand{\psirz}{\psi(\vec{r},0)}
\newcommand{\psirzc}{\psi^*(\vec{r},0)}
\newcommand{\psirtc}{\psi^*(\vec{r},t)}
\newcommand{\probcur}{\vec{S}\ar}
\newcommand{\twodvec}[2]{\left[\begin{array}{cc}{#1} \\ {#2}\end{array}\right]}
\newcommand{\twodmat}[4]{\left[\begin{array}{cc}{#1} & {#2} \\ {#3} & {#4}\end{array}\right]}
\newcommand{\threedvec}[3]{\left[\begin{array}{ccc}{#1} \\ {#2} \\ {#3}\end{array}\right]}
\newcommand{\threedmat}[9]{\left[\begin{array}{ccc}{#1} & {#2}& {#3} \\ {#4} & {#5} & {#6}\\ {#7} & {#8} & {#9}\\\end{array}\right]}
\newcommand{\fourdvec}[4]{\left[\begin{array}{cccc}{#1} \\ {#2} \\ {#3} \\ {#4}\\ \end{array}\right]}
\newcommand{\fourdmat}[4]{\left[\begin{array}{cccc}{#1}\\{#2}\\{#3}\\{#4}\\\end{array}\right]}
\newcommand{\bra}[1]{\langle{#1}|}
\newcommand{\ket}[1]{|{#1}\rangle}
\newcommand{\bkt}[2]{\langle{#1}|{#2}\rangle}
\newcommand{\apdag}{a_+^\dagger}
\newcommand{\amdag}{a_-^\dagger}
\newcommand{\am}{a_-}
\newcommand{\ap}{a_+}
\newcommand{\sol}{\textbf{Solution: }}
\newcommand{\sumn}{ \sum_{n = 0}^N}
\newcommand{\com}[2]{\left[{#1},{#2}\right]}
\newcommand{\h}{\frac{1}{2}}
\newcommand{\hh}{\frac{1}{2}}
\newcommand{\dd}[1]{\mathrm{d}{#1}}
\newcommand{\ddd}[1]{\mathrm{d}^3{#1}}
\newcommand{\ddn}[2]{\mathrm{d^{#1}}{#2}}
\newcommand{\up}{\ket{\uparrow}}
\newcommand{\dn}{\ket{\downarrow}}
\newcommand{\upb}{\bra{\uparrow}}
\newcommand{\dnb}{\bra{\downarrow}}
\newcommand{\vv}[1]{\underline{#1}}
\newcommand{\bigO}{\mathcal{O}}
\newcommand{\del}{\underline \nabla}
\newcommand{\+}[1]{\ensuremath{\mathbf{#1}}}
\newcommand*\rfrac[2]{{}^{#1}\!/_{#2}}

\section{Introduction}
\label{intro}
Exotic dimers consisting of a Rydberg atom bound to a neutral ground state atom possess many fascinating properties, such as their oscillatory potential energy curves, extremely large bond lengths, complex nodal wave functions and, in the polar ``trilobite'' case, huge permanent electric dipole moments ~\cite{GreeneSadeghpourDickinson,HamiltonGreeneSadeghpour,HamiltonThesis,KhuskivadzeChibisovFabrikant, Li}.
In recent years, non-polar long-range Rydberg molecules consisting primarily of $ns$~\cite{Bendkowsky}, $np$~\cite{Sasmannshausen} 
and $nd$ \cite{AndersonRaithelPRL,KruppGaj,GajKrupp} states have been observed in Rb condensates, and a polar trilobite molecule exhibiting a kilo-debye permanent electric dipole moment 
has been photoassociated in Cs ~\cite{Shaffer}. Rydberg molecules have been formed in an optical lattice, providing a non-destructive probe of the Mott transition \cite{Manthey2015}. Recently, molecular formation in non-alkali atomic species have been explored \cite{Eiles, DeSalvo}. 

Current experiments can excite very high Rydberg states in dense condensates so the Rydberg electron's orbit encloses more than one ground state
 atom, increasing the probability of forming polyatomic molecules \cite{Pfau,Bottcher}. At higher densities and excitation energies even the coupling between the Rydberg electron and the entire condensate~\cite{PfauBEC,Karpiuk} can be studied; in this regime the spectrum no longer exhibits few-body molecular lines but rather demonstrates a density shift \cite{GajNatComm} possibly requiring a mix of few and many-body approaches \cite{Schlag,Schmidt2015}. Theoretical 
 efforts in this area have predicted the formation of Borromean trimers~\cite{Rost2009} and investigated the breathing modes of coplanar molecules~\cite{Rost2006}. These investigations have only included $s$-wave scattering, which neglects essential physics, particularly relevant at high density \cite{Pfau,Schlag}. Very recently a Rydberg trimer including $p$-wave scattering and electric field effects has been investigated \cite{newpaper}
 
This present work develops an accurate theoretical framework  incorporating $p$-wave scattering that robustly generalizes to any number of constituent atoms in an arbitrary molecular shape.  General formulas are provided for the electronic wave functions and Born-Oppenheimer adiabatic potential energy curves (APECs) in terms of linear combinations of the diatomic ``trilobite'' \cite{GreeneSadeghpourDickinson} and ``butterfly'' \cite{HamiltonGreeneSadeghpour} wave functions; construction of these hybridized orbitals is aided by adapting them to the molecular symmetry point group using the projection operator method \cite{Bunker, Bunker 2}.  A key result of this study is that the level spacings, degeneracies, and adiabatic/diabatic level crossing properties of these systems are determined by hybridized orbitals reflecting the molecular geometry. This represents a necessary step towards an accurate understanding of experimental spectra in increasingly dense environments.

This work is organized as follows: section 2.1 describes the polyatomic Hamiltonian and outlines the full solution via numerical diagonalization. In section 2.2, the properties of the diatomic molecular states are studied in detail, so that in section 2.3 the polyatomic Schr\"{o}dinger equation can be solved by construction of linear combinations of these diatomic orbitals. In section 2.4, these results are specialized to highly symmetric molecules. Section 3 presents results for a variety of molecular geometries. The reliability and accuracy of the analytic approximations is demonstrated and the effects of symmetry on the molecular spectra and properties are examined in detail. Section 3.4 describes some of the properties of polyatomic molecules formed from low angular momentum Rydberg states, which are relevant to current experimental efforts. Finally, in section 4, the general effects of increasing the number of atoms are studied for the trilobite states, and section 5 concludes with a discussion of experimental proposals for controlled formation of highly symmetric polyatomic molecules.

\section{Theoretical approach}
\label{theory}
\subsection{Pseudopotential and Basis Diagonalization}
\label{diagsec}
The polyatomic system consists of $N$ ground state atoms, located at $\vec R_i=(R_i,\theta_i,\varphi_i)$, 
surrounding a central Rydberg atom. For tractability, the molecular breathing modes, where the ground state atoms share a common distance $R_i = R$ to the Rydberg core, are the primary focus of this study. The $n = 30$ states of rubidium are studied to connect with previous work \cite{GreeneSadeghpourDickinson,Rost2006}, but the general framework applies to any Rydberg state of an alkali with a negative scattering length. Within the Born-Oppenheimer approximation, the nuclei of all atoms are assumed to be stationary with respect to the electronic motion. The Rydberg electron interacts with each neutral atom through the $s$-wave Fermi pseudopotential along with the p-wave scattering term due to Omont \cite{Fermi,Omont}. Since the scale of the Rydberg orbit extends over a far greater range than any interatomic potentials, the Hamiltonian only includes the unperturbed atomic Hamiltonian, $H_0$, and $N$ pseudopotentials:
\begin{align}
\label{ham}
&\hat H_N({\vec r}) = \hat H_0+\sum_{i = 1}^N \hat V(\vec r,\vec R_i)\\
&\nonumber\hat V(\vec r,\vec R_i)= 2\pi\delta({\vec r} - {\vec R_i})\Big(a_s[k(R)] +
3a_p^3[k(R)]{\cev{\nabla}_{ r}}\cdot{\vec {\nabla}_{ r}}\Big).
\end{align}
\noindent
Atomic units are used throughout. The triplet $s$-wave scattering length $a_s$ and $p$-wave scattering volume $a_p^3$ depend on $R$ through the semiclassical relationship $k^2/2 = E_n - 1/R$; the singlet scattering length is an order of magnitude smaller and is ignored. These parameters depend on the energy-dependent triplet scattering phase shifts $\delta_{0}^T$ and $\delta_1^T$ \cite{BahrimThumm} through the relationships
\begin{align}
\begin{split}
\label{scatl}
 & a_{s}[k(R_i)]=-\frac{\tan\left(\delta_{0}^T[k(R)]\right)}{[k(R)]},\\
  & a_{p}^{3}[k(R_i)]=-\frac{\tan\left(\delta_{1}^T[k(R)]\right)}{[k(R)]^{3}}.
\end{split}
 \end{align}
\noindent
The APECs are obtained by diagonalizing equation \eqref{ham} in a basis of hydrogenic eigenstates $\phi_{nlm}(\vec r) = R_{n l}(r)Y_{lm}(\theta,\varphi)$ of the unperturbed Hamiltonian $H_0$,
\begin{equation}
\label{hydwfs}
\left(\hat H_0 +\frac{1}{2(n - \mu_l)^2}\right)\ket{\phi_{nlm}}=0,
\end{equation}
 where $\mu_l$ is the quantum defect for a given angular momentum $l$ relative to the Rydberg core.
Two significant difficulties complicate this approach. Rubidium, along with several other alkali species, possesses a $p$-wave shape resonance that creates an unphysical divergence in the APECs. This is remedied by enlarging the basis size to include additional hydrogenic manifolds adjacent to the manifold of interest, giving sensible results due to level repulsion between states of opposite symmetry \cite{HamiltonGreeneSadeghpour}. This introduces a more serious problem: the delta function potential formally diverges, so the APECs do not converge with the addition of more adjacent manifolds \cite{Fey}, and in fact begin to yield worse results with increasing basis size in comparison with sophisticated Green function techniques that exactly solve the diatomic Hamiltonian.  In the following numerical results the basis consists of the $n$, $n\pm1$ hydrogenic manifolds, which we have found agrees well with the Green function method in the diatomic case.

\subsection{$N = 1$: Diatomic molecular features}
\label{diathe}
\subsubsection{States of low angular momentum:}
\label{diathelowl}
Two distinct classes of APECs, characterized by the unperturbed Rydberg electron's angular momentum, emerge from the diagonalization. The non-zero quantum defects of low-$l$ states separate them energetically from the nearly degenerate high-$l$ ($l \ge 3$ for Rb) manifold. The APECs for low-$l$ states are tens to hundreds of MHz in depth and support only a few weakly bound molecular states. The first order APEC for the $\ket{nlm}$ state including only the $s$-wave interaction is proportional to the Rydberg electron's wave function and, due to the symmetry provided by the internuclear axis, is only non-zero for $m_l = 0$:
\begin{equation}
\label{lowlNone}
 \tilde E_{l\le 2,0}(R) =2\pi a_s[k(R)]\left(\frac{u_{nl}(R)}{R}\right)^2\left|Y_{l0}(0,0)\right|^2.
\end{equation}

\subsubsection{``Trilobite'' and ``butterfly'' states of mixed high angular momentum:}
\label{diathehighl}

In contrast, four strongly perturbed high-$l$ eigenstates form out of the degenerate high-$l$ states: the ``trilobite'' state, dominated by s-wave scattering, and three ``butterfly'' states - ``$R$-butterfly,'' ``$\theta$-butterfly,'' and ``$\varphi$-butterfly - corresponding to the three directional derivatives in the p-wave interaction.  Their APECs are nearly an order of magnitude deeper than the low-$l$ states. As will be seen in the following sections, coherent sums of these diatomic states fully describe the polyatomic states, and so they are studied in detail to clearly elucidate their properties.

 \begin{figure}[h]

 \begin{center}
\includegraphics[scale = 0.4]{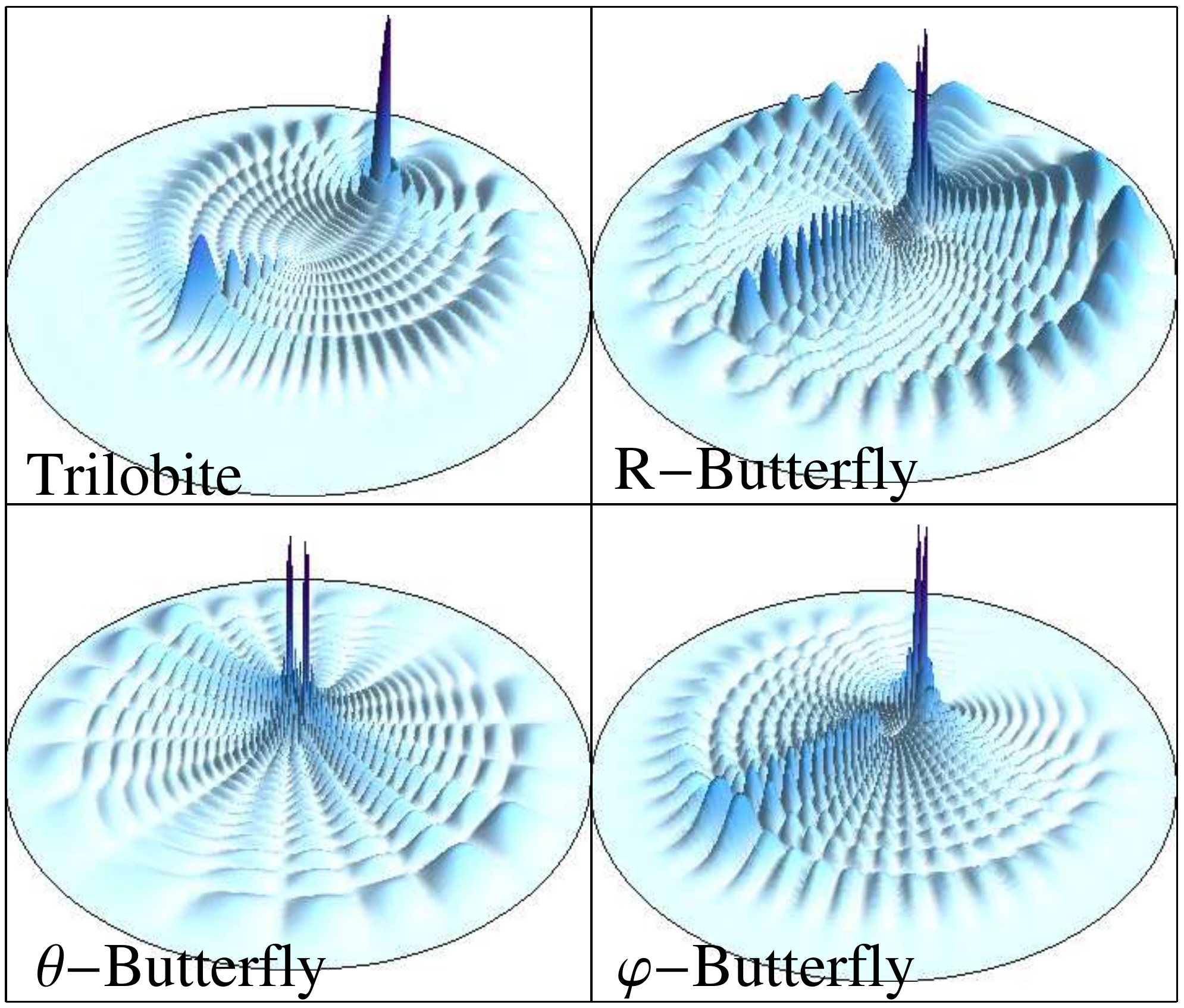}
\caption{Planar cuts of the probability amplitudes $\sqrt{r^2\Psi(x,y,z)^2}$ of the fundamental diatomic eigenstates. The $\theta$-butterfly lies in the $yz$ plane; the three others  in the $xy$ plane. $R=840$ a.u for the trilobite and  $345$ a.u. for the three butterfly states. The Rydberg core is at the center of each figure and the neutral atom is underneath the most prominent spikes.}
 \label{diatomicorbitalsfig}
\end{center}

 \end{figure}

 \noindent
Due to the particular properties of the $s$- and $p$-wave delta-function potentials in equation \eqref{ham}, the diatomic eigenstates, eigenenergies, and  overlap matrix elements for these high-$l$ states can be written \cite{Rost2009} as elements of a ($4\times 4)\otimes (N\times N$) ``trilobite overlap matrix'', $\+{\Upsilon_{pq}^{\pmb\alpha\pmb\beta}}$, where $\left(\+{\Upsilon_{pq}^{\pmb\alpha\pmb\beta}}\right)^*=\+{\Upsilon_{qp}^{\pmb\beta\pmb\alpha}}$. The lower indices $p$ and $q$ label the position vectors $\vec R_p$ and $\vec R_q$ of two neutral atoms; a lower index $r$ indicates $\vec R_r = \vec r$.  Upper indices $\alpha$ and $\beta$ label the eigenstates: the normalized $\xi$th eigenstate associated with an internuclear axis $\vec R$ is given by  $\+{\Upsilon_{Rr}^{\pmb\xi1}}/\sqrt{\+{\Upsilon_{RR}^{\pmb\xi\pmb\xi}}}$.  After defining $a_{\xi = 1} = a_s$ and $a_{\xi \ge 2} = 3a_p^3$ the trilobite ($\xi = 1$) and three ($R$, $\theta$, $\varphi$)-butterfly APECs $(\xi = 2,3,4)$ have the concise form:
 \begin{align}
 \label{trilowfs}
 E_{l>3}^\xi(R)&= 2\pi a_\xi[k(R)]\+{\Upsilon_{RR}^{\pmb\xi\pmb\xi}}.
 \end{align} 
The trilobite overlap matrix is defined as
\be
\label{tomdef}
\+{\Upsilon_{pq}^{\pmb\alpha\pmb\beta}}=\sum_{l=3}^{n-1}\sum_{m = -l}^{m = l}\left[\Phi^\alpha_{nlm}(\vec R_p)\right]^*\Phi^\beta_{nlm}(\vec R_q),
\ee
where the summation extends over energetically degenerate states, starting at $l = 3$ for Rb.  $\Phi^\alpha_{nlm}$ labels the wave function and components of the gradient in spherical coordinates:
\be
\label{estatedef}
\Phi^\alpha_{nlm}(\vec r) =\begin{cases}
\phi_{nlm}(\vec r)&\alpha=1\\ \pde{\phi_{nlm}(\vec r)}{r}&\alpha=2\\
\frac{1}{R}\pde{\phi_{nlm}(\vec r)}{\theta}&\alpha = 3\\
\frac{1}{R\sin\theta}\pde{\phi_{nlm}(\vec r)}{\varphi}&\alpha=4.\end{cases}
 \ee 
 \noindent
If the low-$l$ states are included in equation \eqref{tomdef}, the trilobite eigenstate can be summed analytically using the Green function for the Coulomb problem \cite{ChibisovPRL}
 \begin{align}
  \label{chibisovtrilobite}
&\+{\Upsilon_{Rr}^{11}}= \frac{u_{n0}'(t_-)u_{n0}(t_+) - u_{n0}(t_-)u_{n0}'(t_+)}{4\pi\Delta t},\\
&\Delta t = t_+ - t_-,\nonumber\\&
t_{\pm} =\frac{1}{2}\left(R+r\pm\sqrt{R^2 + r^2 -2Rr\cos\gamma}\right),\nonumber
\end{align}
where $\gamma$ is the angle between $\vec R$ and $\vec r$. This expression is a good approximation for the summation in equation \eqref{tomdef} for energies between the high-$l$ manifold and the low-$l$ states with nonzero quantum defects. 
The three butterfly eigenstates can be found by differentiating equation \eqref{chibisovtrilobite} with respect to  $R$, $\theta_R$, and $\varphi_R$:
\begin{align}
\label{chibisovRbutterfly}
&\+{\Upsilon_{Rr}^{21}} =\frac{(r\cos\gamma-R)\mathcal{F}(t_+,t_-)}{8 \pi\Delta t^3}\\&\,\,\,\,\,\,\,\,\,\,\,\,+\frac{u_{n0}(t_+)u_{n0}''(t_-)-u_{n0}(t_-)u_{n0}''(t_+)}{8 \pi\Delta t}\nonumber\\
\label{chibisovthetabutterfly}
&\+{\Upsilon_{Rr}^{31}}= \cos\theta_R\cos\varphi_R\Upsilon_x + \cos\theta_R\sin\varphi_R\Upsilon_y-\sin\theta_R\Upsilon_z \\
\label{chibisovphibutterfly}
&\+{\Upsilon_{Rr}^{41}}= -\sin\varphi_R\Upsilon_x + \cos\varphi_R\Upsilon_y,\end{align}
where 
\begin{align}
&\Upsilon_{x,y,z} = \frac{\mathcal{F}(t_+,t_-)}{8\pi(\Delta t)^3}(\sin\theta_r\cos\varphi_r,\sin\theta_r\sin\varphi_r,\cos\theta_r),\nonumber\\
\fl&\mathcal{F}(t_+,t_-)=- 2(\Delta t)  u_{n0}'(t_+) u_{n0}'(t_-)\nonumber\\&\,\,\,\,\,\,\,\,-u_{n0}(t_-)[2 u_{n0}'(t_+) - (\Delta t) u_{n0}''(t_+)]\nonumber \\&\,\,\,\,\,\,\,\, +  u_{n0}(t_+)[ 2u_{n0}'(t_-) +(\Delta t)  u_{n0}''(t_-)].\nonumber
\end{align}
\noindent
The $\theta,\varphi$ butterfly orbitals can be identified as vectors of magnitude $\frac{\mathcal{F}(t_+,t_-)}{8\pi(\Delta t)^3}$ parallel to the $\theta,\varphi$ unit vectors; the trilobite and $R$-butterfly orbitals are fully symmetric about the internuclear axis. The diagonal elements $\+{\Upsilon_{pp}^{\xi\xi}}$ are obtained by evaluating equations (\ref{chibisovtrilobite} - \ref{chibisovphibutterfly}) in the limit $\vec R_p\to \vec R_q$:
\begin{align}
\label{diagonalTrilo}
&\+{\Upsilon_{RR}^{11}}= \frac{(2n^2 - R)\left(u_{n0}(R)/n\right)^2+ Ru_{n0}'(R)^2}{4\pi R}\\
\label{diagonalRbutterfly}
&\+{\Upsilon_{RR}^{22}}= \+{\Upsilon_{RR}^{33}}-\frac{u_{n0}(R)}{12\pi R^3}\left[3Ru'_{n0}(R)+2u_{n0}(R)\right]
\end{align}
\begin{align}
\label{diagonalAngbutterfly}
&\+{\Upsilon_{RR}^{33}}=\+{\Upsilon_{RR}^{44}}\\
&\,\,\,\,\,\,\,\,\,\,\,\,\,\,= \frac{4\pi R(2n^2 - R)\+{\Upsilon_{RR}^{11}} - n^2u_{n0}'(R)u_{n0}(R) }{12\pi n^2R^2}.\nonumber
\end{align}
The diatomic angular butterfly APECs, $\+{\Upsilon_{RR}^{33}}$ and $\+{\Upsilon_{RR}^{44}}$,  are degenerate $^3\Pi$ molecular states and, in contrast to the $^3\Sigma$ trilobite or $R$-butterfly APECs, do not oscillate as a function of $R$. 
This is due to destructive interference between terms in the numerator of equation (\ref{diagonalAngbutterfly}).

The full Hamiltonian is solved by assuming a linear combination of the eigenstates in equation \eqref{tomdef} as a solution \cite{KurzThesis}. Two additional properties of the overlap matrix are important.  $\+{\Upsilon_{pq}^{\pmb\alpha\pmb\beta}}$ is the overlap between different diatomic orbitals $\alpha$ and $\beta$ associated with different ground state atoms located at $\vec R_p$ and $\vec R_q$, respectively, and the matrix element of the $\xi$th interaction term of the Hamiltonian between an orbital $\alpha$ located at $\vec R_p$ and an orbital $\beta$ located at $\vec R_q$ is $\+{\Upsilon_{ip}^{\pmb\xi\pmb\alpha}}\+{\Upsilon_{qi}^{\pmb\beta\pmb\xi}}$.  A generalized eigenvalue equation for $E(R)$ is then obtained:
 \be
 \label{highlNone}
\sum_{\beta=1}^4 \left(\sum_{\xi = 1}^4a_\xi(k)\+{\Upsilon_{RR}^{\pmb\xi\pmb\alpha}}\+{\Upsilon_{RR}^{\pmb\beta\pmb\xi}}-\frac{E}{2\pi}\+{\Upsilon_{RR}^{\pmb\alpha\pmb\beta}}\right) \Omega^\beta_R = 0.
 \ee
Throughout the rest of this paper the explicit dependence of $k$, $E$, and the eigenvector $\vec\Omega$ on the internuclear distance $R$ is assumed for brevity.

\subsection{$N= $ many: Generalization to polyatomic molecular states}
\label{polythe}
The diatomic results described above readily generalize to the $N>1$ case. For the low-$l$ APECs, all $m_l$ values are allowed; therefore $N_d=2l+1$ degenerate $m_l$ states mix together. This causes $N_p = \max(4N,N_d)$ APECs to split away from the unperturbed electronic states.   For the $s$-wave interaction alone the APECs are the eigenvalues $E(R)$ given by the matrix equation
\begin{align}
\label{lowlNmany}
&a_s(k)\left(\frac{u_{nl}(R)}{R}\right)^2\sum_{m=-l}^{l}\sum_{i=1}^NY_{lm'}^*(\theta_i,\varphi_i)Y_{lm}(\theta_i,\varphi_i)\Omega_m^1\nonumber\\&=\frac{E}{2\pi }\Omega_{m'}^1.
\end{align}
The trilobite overlap matrix formalism allows for rapid generalization to the polyatomic system, since the trial solution used to obtain equation \eqref{highlNone} is expanded to include linear combinations of trilobite and butterfly eigenstates for each diatomic Rydberg-neutral pair:
\be
\label{lctrilobutterfly}
\Psi(\vec r) = \sum_{p= 1}^N\sum_{\alpha = 1}^4\Omega_{p}^\alpha\+{\Upsilon_{pr}^{\pmb\alpha 1}}.
\ee
This formulation provides key physical insight and also greatly reduces the calculational effort to the diagonalization of at most a $4N\times4N$ matrix, rather than the full $n^2\times n^2$ basis size needed to diagonalize equation (\ref{ham}). The trilobite APECs, only including the $s$-wave interaction, are the eigenvalues $E(R)$ satisfying
\be
\label{highlNmanyswave}
\sum_{q = 1}^N\left( a_s(k)\+{\Upsilon_{pq}^{11}}-\frac{E}{2\pi}\+{\pmb\delta_{pq}}\right)\Omega_q^1 =0.
\ee 
The p-wave interaction is included analogously to the diatomic case, yielding a generalized eigenvalue problem with a $4N\times 4N$ matrix:
\begin{align}
\label{highlNmanyspwave}
\sum_{\beta=1}^4\sum_{p,q=1}^N\Bigg(\sum_{\xi = 1}^4&\sum_{i=1}^Na_\xi(k)\+{\Upsilon_{ip}^{\pmb\xi\pmb\alpha}}\+{\Upsilon_{qi}^{\pmb\beta\pmb\xi}}-\frac{E}{2\pi}\+{\Upsilon_{pq}^{\pmb\alpha\pmb\beta}}\Bigg)\Omega_{q}^{\beta}=0.
\end{align}
Equations (\ref{lowlNmany}-\ref{highlNmanyspwave}) accurately reproduce the full diagonalization results for arbitrary molecular configurations and numbers of atoms, particularly for the trilobite states.  Due to the challenges with basis set diagonalization described in section \eqref{diagsec}, spectroscopic accuracy for the low-$l$ states can only be achieved through a careful convergence study \cite{Schlag} or with the Green function method. As a result, equations \eqref{lowlNone} and  \eqref{lowlNmany} should only be used for qualitative study. For the $n = 30$ system studied here, the $l = 0$ states are about 20\% deeper than these first-order results predict. In contrast, the high-$l$ APECs are quite accurate. Since only one manifold is included, these formulas break down at internuclear distances smaller than the location of the $p$-wave shape resonance, and also do not describe couplings with low$-l$ states at large detunings. For investigations in regimes where these inaccuracies are irrelevant, equation \eqref{highlNmanyspwave} is a valuable computational advance, particularly for experiments probing high Rydberg states up to $n \sim 110$ \cite{Karpiuk,Schlag} due to the reduced basis size.

The off-diagonal elements of $\+{\Upsilon_{pq}^{\pmb\alpha\pmb\beta}}$, corresponding to the overlap between orbitals associated with different Rydberg-neutral pairs, determine the size of the differences between the polyatomic states and the $N = 1$ state. In the absence of these overlaps equation \eqref{highlNmanyspwave} is diagonal in the lower indices and all $N$ polyatomic APECs converge to the diatomic APEC.  At large $R$ the overlap between orbitals vanishes, and the APECs are seen to converge to the diatomic limit. Co-planar molecules typically exhibit larger splittings than three-dimensional molecules for this same reason. Additionally, as $N$ grows the system will deviate more strongly from the $N = 1$ case; this causes the global minimum of the APECs to deepen with $N$. The angular dependence of the trilobite wave functions contributes considerably to the energy landscape of the system at hand, especially when two ground state atoms are close in proximity and therefore have a large overlap. More stable Rydberg molecules can thus be engineered by exploiting these features.

\begin{figure}[h]
\begin{center}
\subfigure[]{\label{octagonlayout}\includegraphics[scale = 0.25]{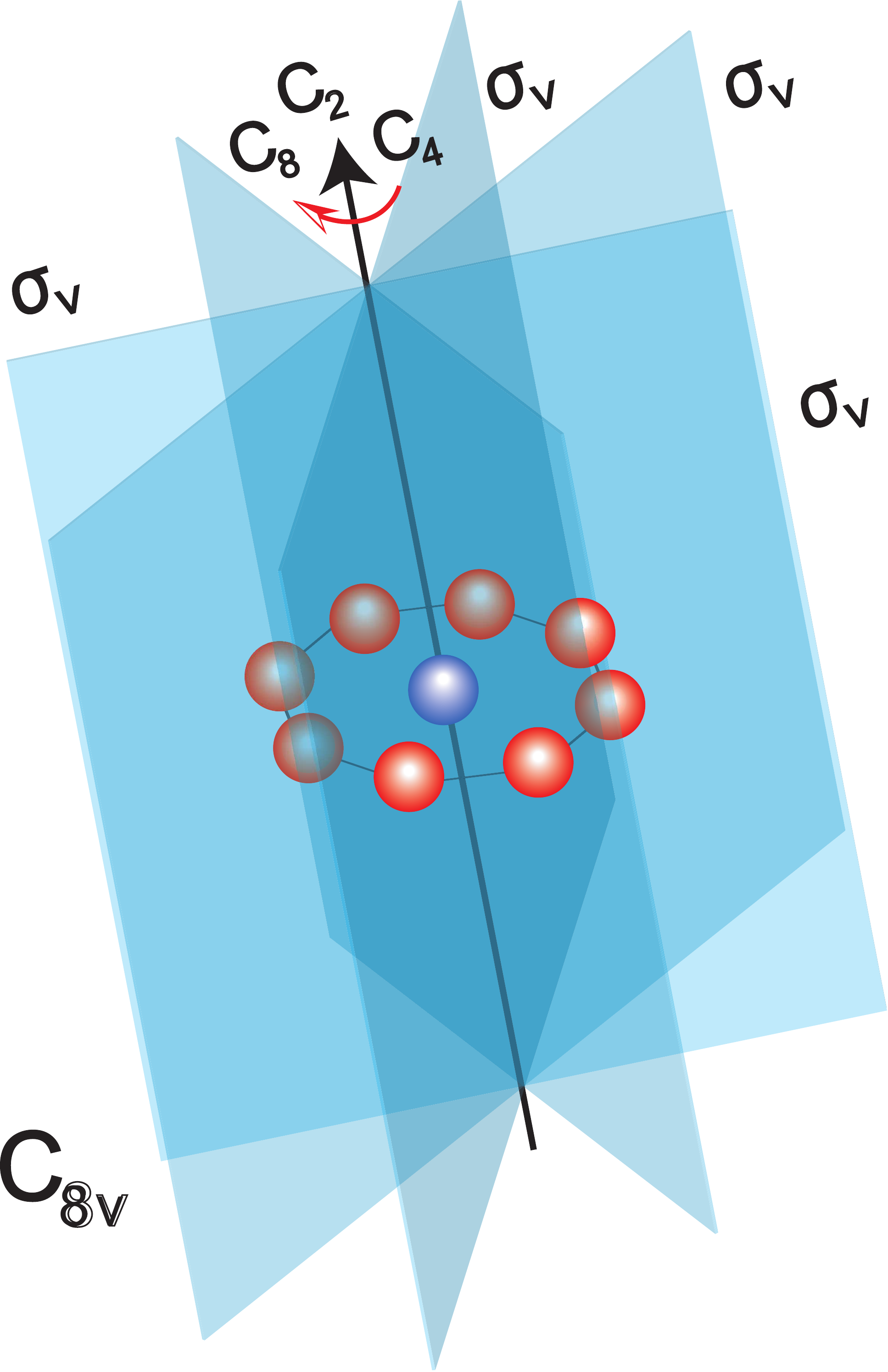}}\\\subfigure[]{\label{hoodoo2}\includegraphics[scale = 0.4]{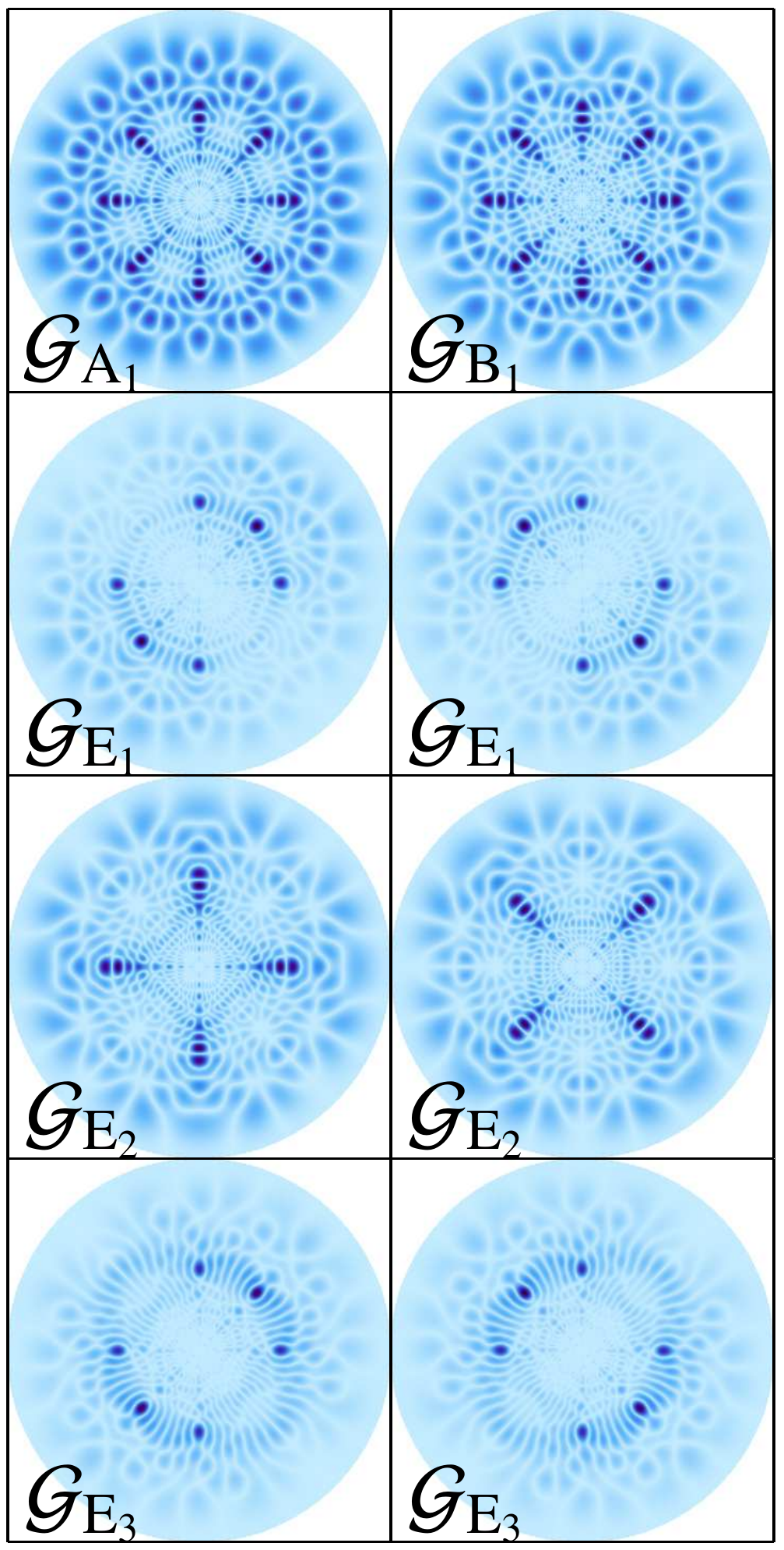}}\subfigure[]{\label{hoodoo3}\includegraphics[scale = 0.4]{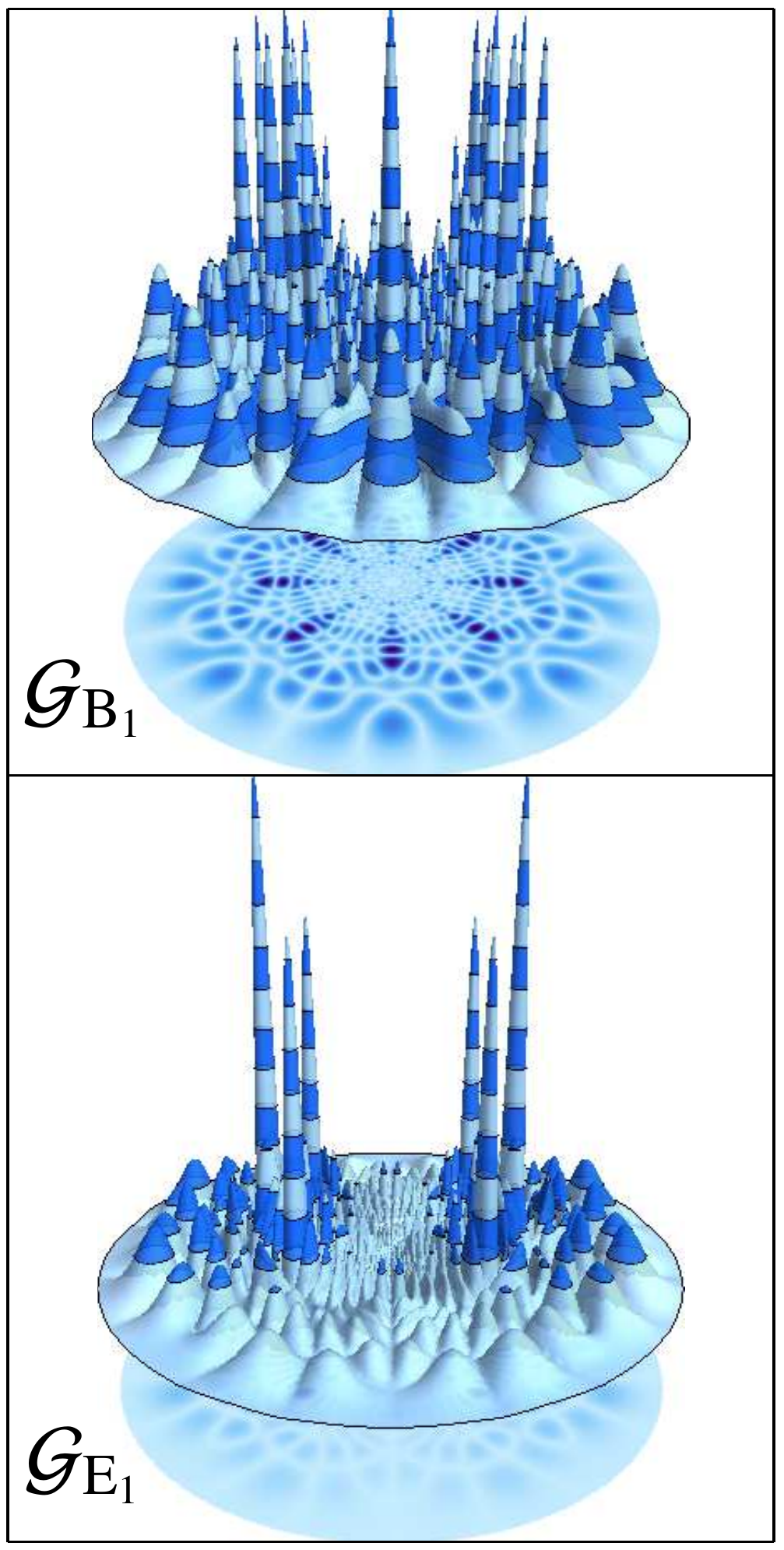}}
\end{center}
\begin{center}
\caption{(a) The symmetry operations for $C_{8v}$ symmetry. The $\sigma_d$ reflection planes bisect the lines between ground state atoms and are not shown for clarity. (b) ``Hoodoo'' symmetry adapted orbitals for trilobite states of an octagonal molecule with internuclear distance $R = 840$ a.u. The probability amplitude $\sqrt{r^2|\psi(x,y,0)|^2}$ is plotted in the $xy$ plane. (c) The  electron probability corresponding to the one-dimensional irrep $B_1$ (top) and one of the doubly-degenerate $E_1$ irreps (bottom) are plotted. }
\end{center}
\label{mainhoodoofigure}
\end{figure}

\subsection{Symmetry-adapted orbitals}
\label{symthe}
\subsubsection{Molecular symmetry point groups.} To fully understand the structure of these APECs, in particular the appearance of degeneracies and level crossings in highly symmetric molecular geometries and the effects of the molecular symmetry on the coupling between trilobite and butterfly states, it is mandatory to characterize the symmetry group of the molecule. The molecular symmetry group is a 
subgroup of the complete nuclear permutation inversion group of 
the molecule~\cite{Bunker, Bunker 2}, which commutes with the molecular 
Hamiltonian in free space. Therefore, the eigenstates of such a 
Hamiltonian can be classified in terms of the irreducible 
representations (irreps) of the given molecular symmetry group, called {\it symmetry-adapted orbitals} (SAOs). Given a molecular symmetry 
group, it is possible to calculate the SAOs associated 
with each irrep of the group using the projection operator method, where the projection operator is \cite{Bunker}
\begin{equation}
\label{projdef}
\hat{\mathcal{P}}^{j}=\frac{l_{j}}{h}\sum_{i=1}^h\chi_{ji}\mathcal{R}_{i}. 
\end{equation}
\noindent
The index $j$ labels the different irreps and 
$i$ denotes the group elements. These are the familiar symmetry operations: rotations, 
reflections, and inversions. $\mathcal{R}_{i}$ represents the operator associated 
with the $i$th symmetry operation; $l_{j}$ and 
$\chi_{ji}$ represent the dimension and character for the $i$th operation, respectively. Finally, $h$ stands for the order of the 
group. The trace of the projection operator, $\tr\hat{\mathcal{P}}^{j}=l_j$, determines the decomposition of the point group into irreps. All irreps with $l_j\ne 0$ are contained.   SAOs associated with different irreps have different parity under the molecular symmetry group, and hence will exhibit real crossings. $l_j$ determines the degeneracy of each irrep.

The projection operator also gives the coefficients $\mathcal{A}_p^{(\alpha,j)}$ for the SAO $\mathcal{G}^{(\alpha,j)}(\vec r)$ corresponding to the $\alpha$th orbital and $j$th irrep:
\be
\label{sao}
\mathcal{G}^{(\alpha,j)}(\vec r)=\sum_{p = 1}^N\+{\Upsilon_{pr}^{\pmb\alpha\pmb1}}\mathcal{A}_p^{(\alpha,j)}.
 \ee
The prescription for calculating the projection operator depends on the orbital in question. The $^3\Sigma$ trilobite and $R$-butterfly states can be symmetry-adapted independently since they are non-degenerate. Since these orbitals are symmetric about the Rydberg-neutral internuclear axis the symmetry operations leave the orbitals unchanged except for an overall transformation of the atomic positions within the molecule, i.e. a permutation of the basis of Rydberg-neutral pairs at different positions $\psi_p$: $\vec\Psi = (\psi_A,\psi_B,...,\psi_N)^T$. The $N\times N$ matrix representations of the symmetry operations can then be identified with a modicum of effort and the sum in \eqref{projdef} performed. The orthogonalized rows of $\hat{\mathcal{P}}^j$ provide the coefficients. 

Since the $\theta$ and $\varphi$ butterfly $^3\Pi$ states are degenerate, these orbitals can be mixed by symmetry operations, so these orbitals must be symmetry-adapted together. The basis size is doubled to allow mixing: $\vec\Psi = (\psi^\theta_A,\psi^\theta_B,...,\psi^\theta_N,\psi^\varphi_A,\psi^\varphi_B,...,\psi^\varphi_N)^T$. The effect of a symmetry operation on the entire molecule transforms orbitals located at one Rydberg-neutral pair to another as in the trilobite/$R$-butterfly case: $\psi^\theta_i\to\psi^\theta_{i'}$ and $\psi^\varphi_i\to\psi^\varphi_{i'}$. However, the symmetry operation now modifies the orbitals themselves. The angular butterfly orbitals are vectors in Cartesian coordinates (see equations (\ref{chibisovthetabutterfly}-\ref{chibisovphibutterfly})) and the symmetry operators in the $xyz$ coordinate basis affect them. This transforms $\psi^\theta_i\to \alpha\psi^\theta_{i'} + \beta\psi^\varphi_{i'}$ and $\psi^\varphi_i\to\gamma\psi^\theta_{i'} + \delta\psi^\varphi_{i'}$; the coefficients $\alpha,\beta,\gamma,\delta$ must then be solved to identify the matrix representation of that symmetry operation.  An explicit example of this process is shown in \ref{appendo}; the final result is the full tabulation of the irreps corresponding to each orbital and the sets of coefficients $\mathcal{A}_p^{(\alpha,j)}$  providing the correct SAOs. These coefficients are listed in \ref{appendo} for the molecular symmetries exemplified in section 3. 
\subsubsection{APECs with symmetry-adapted orbitals.}
The trilobite and $R$-butterfly orbitals always belong to the same irreps as they have identical decompositions, while the angular butterflies have different decompositions that may still share some irreps with the trilobite. As a result each of these possible cases requires a slightly different calculation: the APECs are solutions to a generalized eigenvalue problem for a matrix of $1\times 1$ to $3\times 3$ dimension. These expressions are listed below, starting first with the trilobite APECs to allow for comparison with previous work.
\newline
{\it Trilobite:}
The trilobite APEC for the $j$th irrep satisfies the particularly elegant expression
\be
\label{saoswave}
E^{(j)} = 2\pi a_s(k)\sum_{p,q=1}^N\mathcal{A}_p^{(1,j)}\+{\Upsilon_{pq}^{11}}\mathcal{A}_q^{(1,j)}.
\ee
This equation exactly reproduces the results calculated more laboriously in \cite{Rost2006}. 
In the following equations the explicit dependence $j$ is dropped for brevity.

\noindent
 {\it Trilobite and R-butterfly: } These APECs are given by the generalized eigenvalues of the $2\times2$ matrix equation
\begin{align}
\label{saospwavetriloR}
\sum_{\beta=a}^b\sum_{p,q=1}^N&\mathcal{A}_p^{\alpha}\mathcal{A}_q^{\beta}\left(\sum_{i=1}^N\nonumber\sum_{\xi = a}^ba_\xi(k)\+{\Upsilon_{ip}^{\pmb\xi\pmb\alpha}}\+{\Upsilon_{qi}^{\pmb\beta \pmb\xi}}-\frac{E}{2\pi}\+{\Upsilon_{pq}^{\alpha\beta}} \right)\Omega^{(\beta)}\\&=0;\,\,a=1,b=2.
\end{align}
 {\it Angular butterflies: } The angular butterfly APEC is the solution $E$ to equation \eqref{saospwavetriloR} after setting $a = 3$, $b = 4$ and summing over $\alpha$ from $a$ to $b$. 
 
 \noindent
 {\it All orbitals:}
 When all orbitals correspond to the same irrep, the APECs are given by a $3\times 3$ generalized eigenvalue problem
\begin{align}
\label{saospwave}
\sum_{\beta=1}^3\sum_{p,q=1}^N\left(\sum_{i = 1}^N\sum_{\xi = 1}^4a_\xi(k)\underline{\+{\Upsilon_{ip}^{\pmb\xi\pmb\alpha}}}\widetilde{\+{\Upsilon_{qi}^{\pmb\beta \pmb\xi}}}-\frac{E}{2\pi}\+{\Psi_{pq}^{\pmb\alpha\pmb\beta}}\right)\Omega^{(\beta)}=0,
\end{align}
where the following terms have been defined to incorporate the simultaneously symmetry-adapted $\theta$ and $\varphi$ butterflies by adding the symmetry-adapted $\varphi$-butterfly ($\alpha,\beta = 4$) orbital to the symmetry-adapted $\theta$-butterfly ($\alpha,\beta = 3$) orbital.
\begin{align}
\underline{\+{\Upsilon_{qp}^{\pmb\alpha\pmb\beta}}}&= \left[\+{\Upsilon_{qp}^{\pmb\alpha\pmb\beta}}\mathcal{A}_q^{(j,\alpha)}+\delta_{\beta,3}\+{\Upsilon_{qp}^{\pmb4 \pmb\alpha}}\mathcal{A}_q^{(j,4)}\right]\nonumber\\
\widetilde{\+{\Upsilon_{qp}^{\pmb\alpha\pmb\beta}}}&=\left[\+{\Upsilon_{qp}^{\pmb\alpha\pmb\beta  }}\mathcal{A}_q^{(j,\alpha)}+\delta_{\alpha,3}\+{\Upsilon_{qp}^{\pmb4 \pmb\beta}}\mathcal{A}_q^{(j,4)}\right]\nonumber\\
\+{\Psi_{pq}^{\pmb\alpha\pmb\beta}}&= \mathcal{A}_p^{(j,\alpha)}\+{\Upsilon_{pq}^{\pmb\alpha\pmb\beta}}\mathcal{A}_q^{(j,\beta)}+\delta_{\beta3} \mathcal{A}_p^{(j,\alpha)}\+{\Upsilon_{pq}^{\pmb\alpha4}}\mathcal{A}_q^{(j,4)}\nonumber\\&+\delta_{\alpha3} \mathcal{A}_p^{(j,4)}\+{\Upsilon_{pq}^{4\pmb\beta}}\mathcal{A}_q^{(j,\beta)} + \delta_{\alpha3}\delta_{\beta3} \mathcal{A}_p^{(j,4)}\+{\Upsilon_{pq}^{44}}\mathcal{A}_q^{(j,4)},\nonumber
\end{align}
where $\delta_{mn}$ is the Kronecker delta.

\section{Results}
\label{results}
The APECs of a co-planar octagonal molecule and a body-centered cubic molecule are presented to demonstrate the accuracy of this general formulation. 
 
\subsection{Co-planar geometry: octagonal configuration}
\label{octagon}

The molecular symmetry group of the octagonal configuration is the point group $C_{8v}$, 
depicted  in figure \eqref{octagonlayout}. The reducible representation $\Gamma_{C_{8_v}}$ decomposes into seven total irreps,
\begin{equation}
\label{C8v}
\Gamma_{C_{8_v}}= A_1 \oplus B_1 \oplus E_1 \oplus E_2 \oplus E_3,
\end{equation}
 for the trilobite, $R$-butterfly, and $\theta$-butterfly orbitals, and 
 \begin{equation}
\label{C8vphi}
\Gamma_{C_{8_v}}^\varphi= A_2 \oplus B_2 \oplus E_1 \oplus E_2 \oplus E_3
\end{equation} for the $\varphi$-butterfly orbital. 
The $\theta$-butterfly orbital is completely decoupled from the rest due to its node in the molecular plane; this is a general feature of coplanar molecules. The $\varphi$-butterfly has a different decomposition than the others because of its particular symmetry properties, as discussed in section 2. It is therefore decoupled for the one-dimensional symmetries but couples with the trilobite and $R$-butterfly orbitals for the doubly-degenerate symmetries, resulting in avoided crossings between these APECs.

Symmetry adapted orbitals for the trilobite APECs are displayed in figures \eqref{hoodoo2} and \eqref{hoodoo3}. The internuclear distance is $R = 840$ a.u., the location of the deepest potential well in figure (\ref{trilobiteplots}(a)). These ``hoodoo" states, nicknamed for their
resemblance to the geological formations commonly found in the American Southwest, explicitly exhibit the allowed symmetries. The beautiful nodal patterns in these curves are the result of interference between trilobite orbitals, which as discussed in section \eqref{polythe} is a clear signature of deviations from the $N = 1$ APEC.

\begin{figure}[b]
\begin{center}
\subfigure[]{\label{randomocto1}\includegraphics[scale = 0.4]{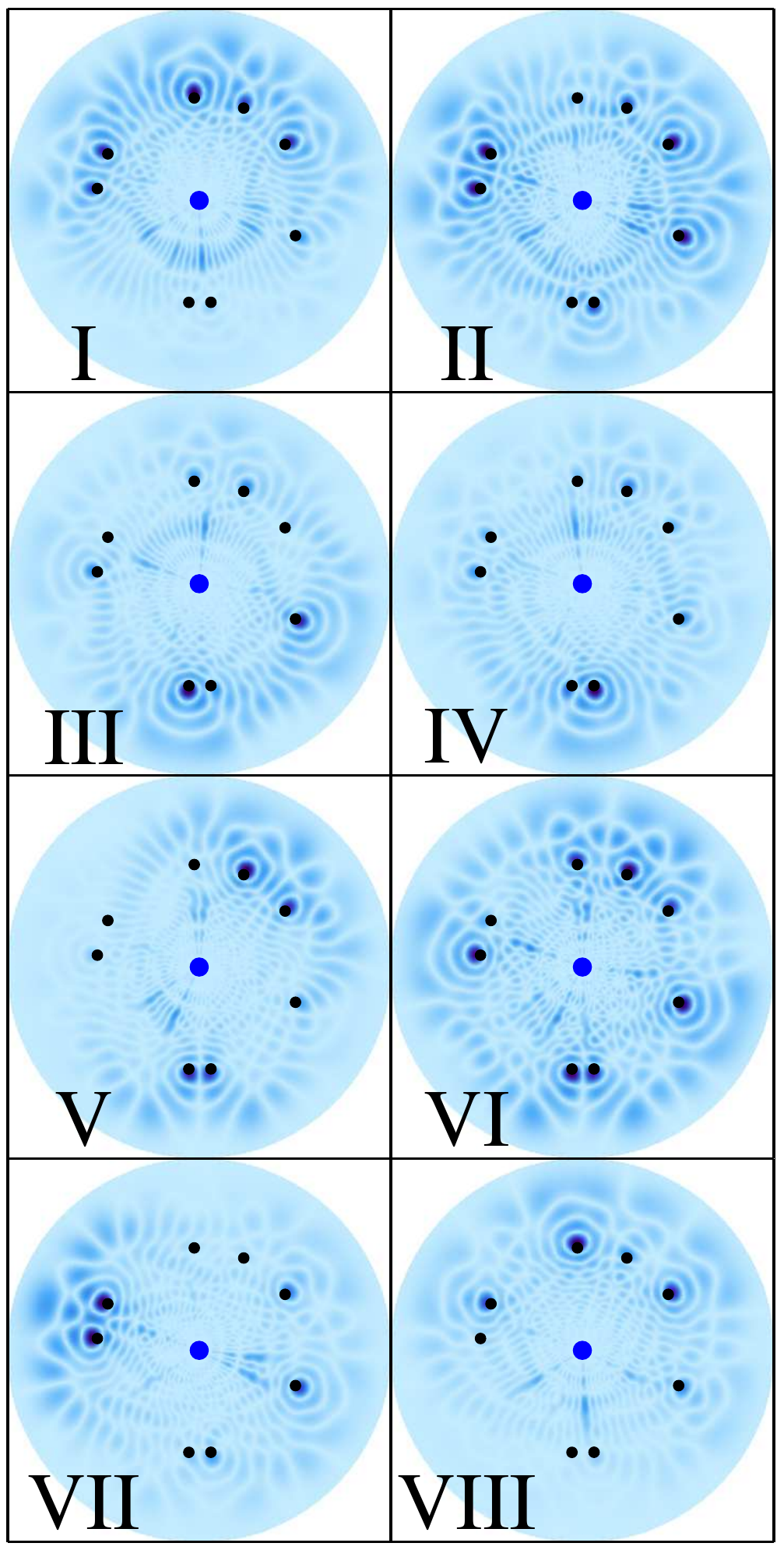}}
\subfigure[]{\label{randomocto2}\includegraphics[scale = 0.4]{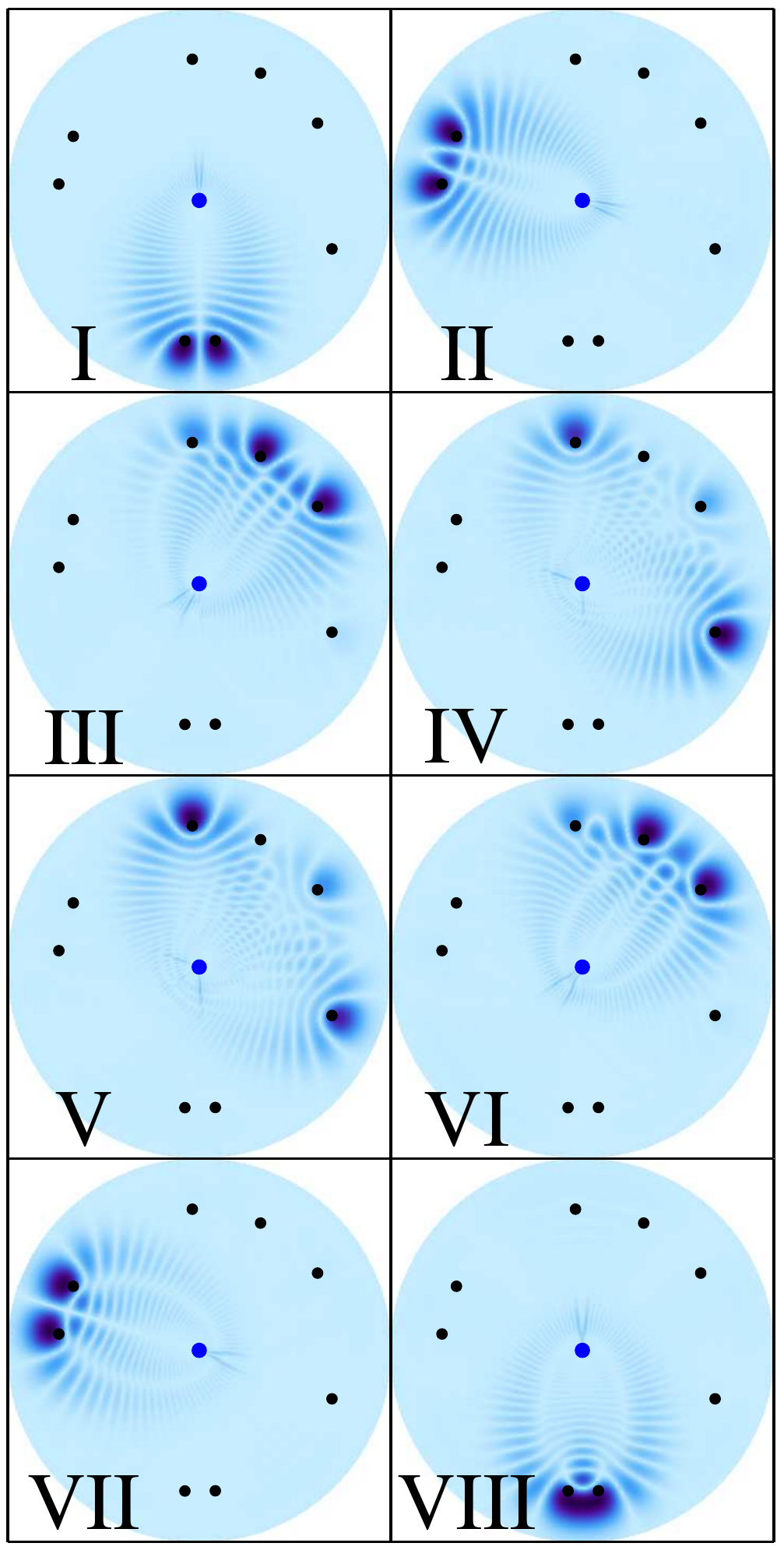}}
\caption{(a) Hybridized trilobite orbitals for a randomly oriented molecular configuration. The Rydberg atom is located at the blue point in the middle of each panel and the neutral atoms are placed at the black points. The common distance $R$ is 1115 a.u. at a local minimum in the lowest energy potential curve. The eigenstates depicted here correspond to APECs that increase in energy from left to right, top to bottom. 
(b) The same as (a), but in the well the farthest distance from the Rydberg atom at $R = 1537$ a.u. }
\label{randomoctagon}
\end{center}
\end{figure}

The full APEC results are shown in figure \eqref{trilobiteplots}, where the exact full diagonalization (black lines) and symmetry-adapted orbital calculation from equations (\ref{saospwavetriloR}-\ref{saospwave}) (colored points) are compared. The dispersion between  different irreps is clearly observed, as has been previously predicted 
 for smaller systems.~\cite{Rost2006}. Excellent agreement between the exact and the symmetry-adapted orbital approach is apparent.

 \subsection{Two-dimensional geometry: random configuration}
 \label{random}
  
Molecules that are not configured symmetrically can still be studied via equation \eqref{highlNmanyspwave}, and the contrasts between these results and those of highly symmetric configurations are of substantial interest. As an example, the hybridized trilobite orbitals for a co-planar, randomly structured geometry at two Rydberg-neutral internuclear distances are displayed in figure \eqref{randomoctagon}. The orbitals at the smaller internuclear distance show substantial interference patterns. Interestingly, for each APEC the electron probability tends to be localized on a subset of the neutral atoms. This subset varies between APECs and is especially clear at the larger internuclear distance displayed in (\ref{randomoctagon}b). A possible explanation stems from semi-classical periodic orbit theory, since the trilobite state $\+{\Upsilon_{Rr}^{11}}$ forms due to interference between the four semiclassical elliptical trajectories focused on the Rydberg core and intersecting at both the neutral atom and at the observation point $\vec r$ \cite{Granger}. Since an ellipse focused on the Rydberg core can lie on at most two neutral atoms, this mechanism is a plausible explanation for why these hybridized orbitals tend to be most localized on two neutral atoms. This phenomenon is not seen in highly symmetric molecules, like the octagon of figure \eqref{mainhoodoofigure}, since the atoms bound by the Rydberg orbit are here determined by the irrep. An additional feature of these unstructured molecules is that deepest/shallowest well partners, i.e. I and VIII, II and VII, etc. are localized on the same atoms but possess different parity with respect to reflection through the $xy$ plane.
\begin{figure}[t]
\begin{center}
\includegraphics[scale = 0.4]{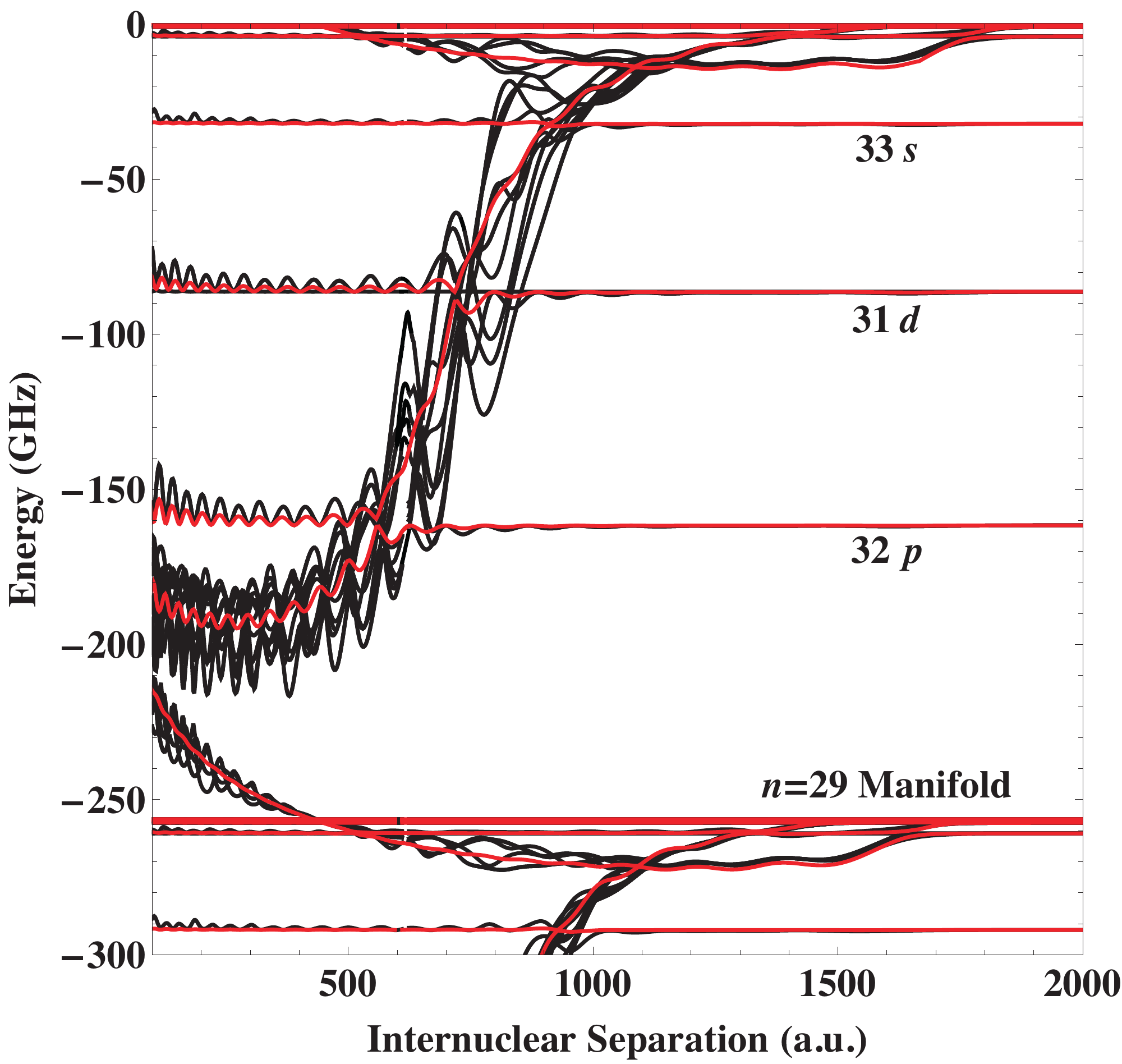}
\caption{(Color online) APECs for Rydberg states between the $n = 29$ and $30$ manifolds for the cubic molecular geometry. The diatomic potential curves are plotted for comparison in red. Large potential wells deep in the butterfly potential curves form, in stark contrast to the diatomic case.}
\label{cubepwavefull}
\end{center}
\end{figure}

\begin{figure}[b]
\begin{center}
{\label{octagon3dcuts}\includegraphics[scale = 0.5]{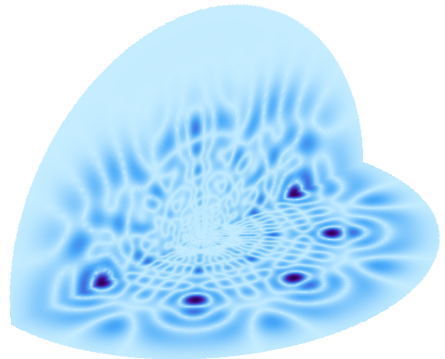}}{\label{cube3dcuts}\includegraphics[scale = 0.5]{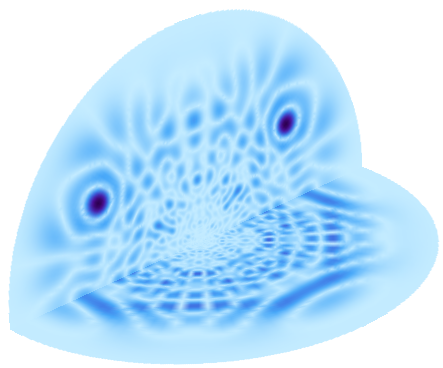}}\\

\includegraphics[scale = 0.4]{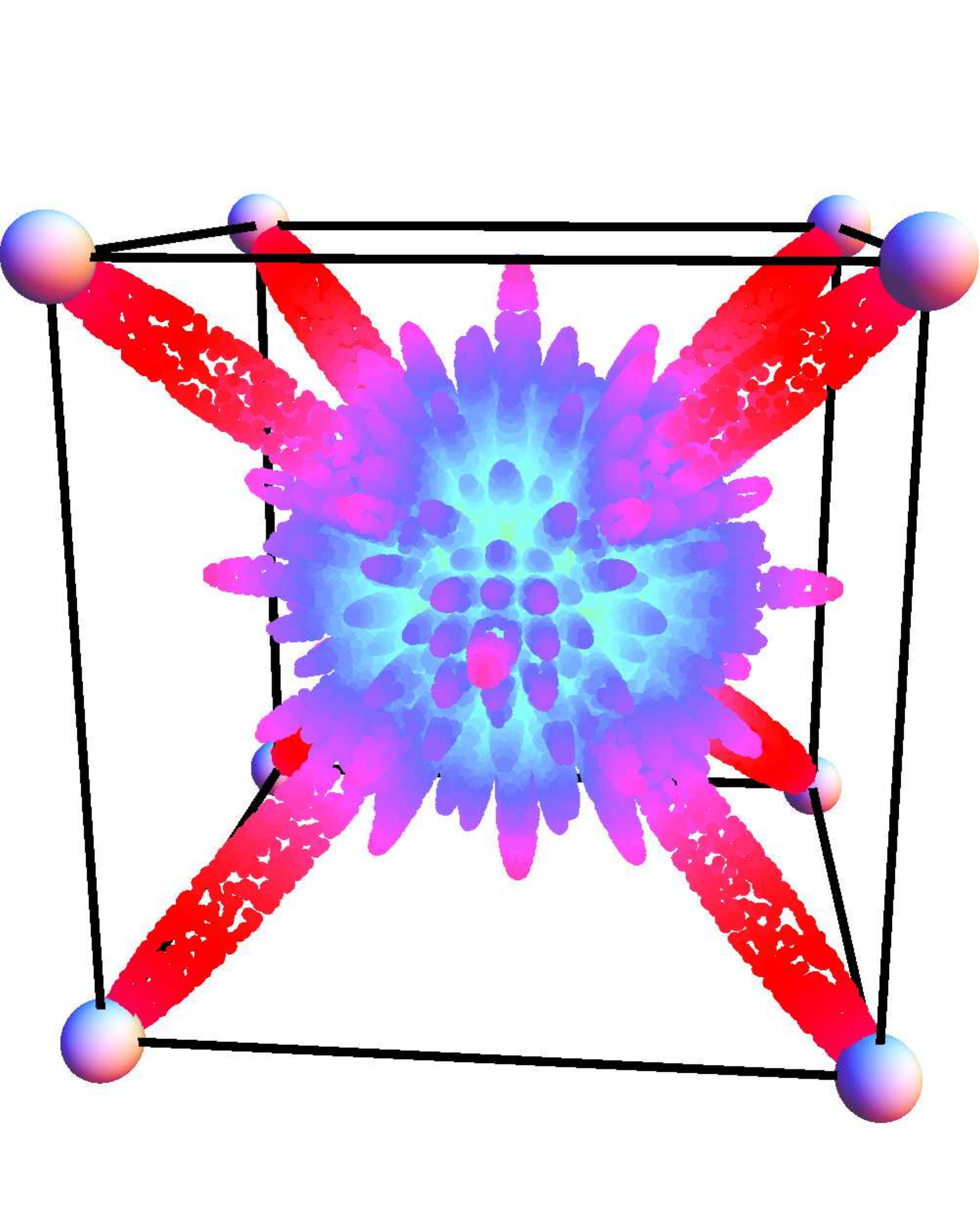}
\vspace{-30pt}
\end{center}
\begin{center}
\caption{Probability amplitudes cuts in the $xy$ and $yz$ planes for the $A_{1g}$ symmetry species of the trilobite-dominated hybridized orbitals, showing the intricate interference patterns for (a) the octagon and (b) the cube (the $yz$ plane here passes through the diagonal of a face). 
(c) A polar plot of the $A_{1g}$ cubic hybridized trilobite state: the probability amplitude, scaled to highlight the nodal structure, is plotted as the distance from the center as a function of $\theta$ and $\varphi$. }
\label{3dcuts}
\end{center}
\end{figure}

\subsection{Three-dimensional geometry: cubic and asymmetric molecules.}
\label{cube}

\begin{figure*}[t]
\begin{center}
\includegraphics[scale = 0.45]{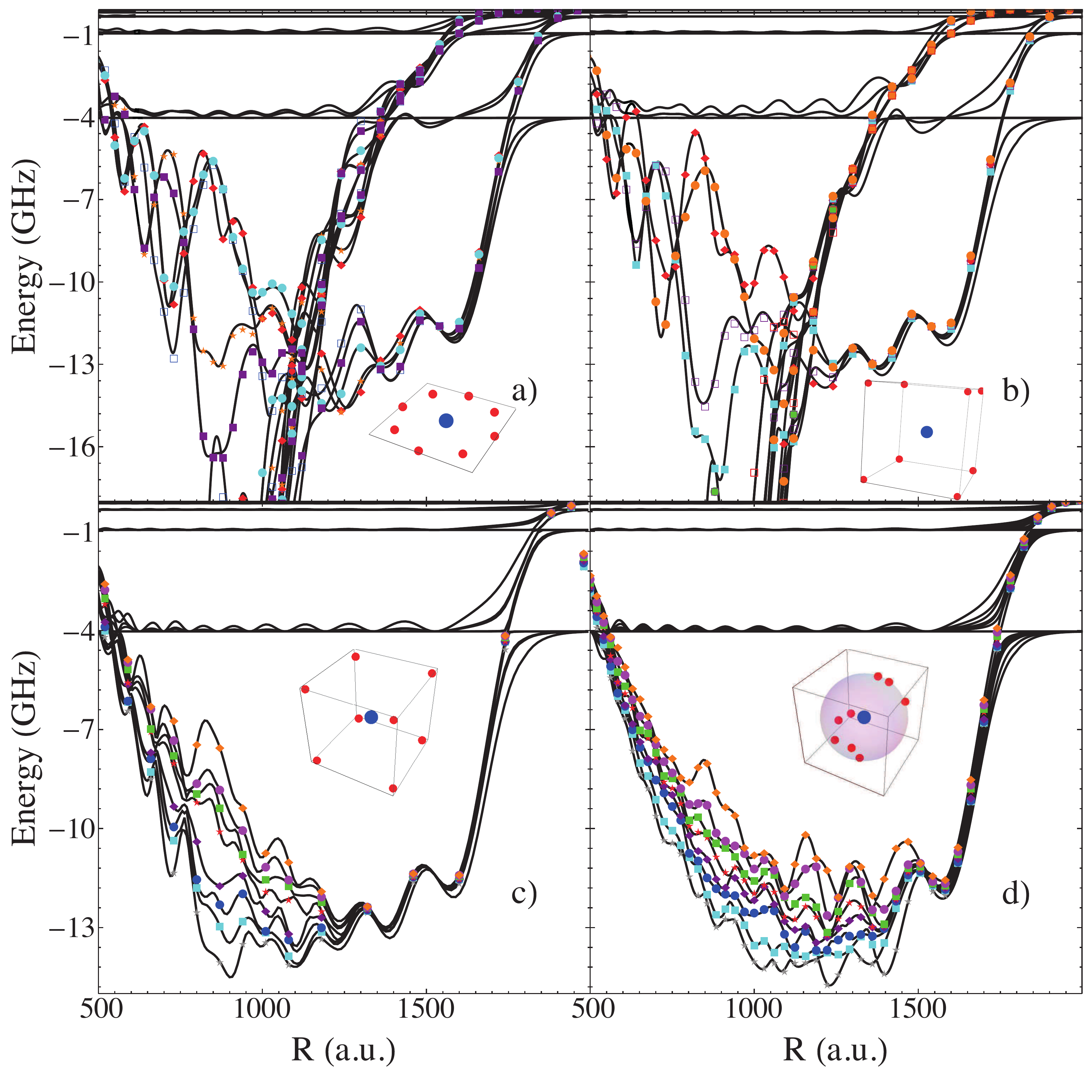}
\caption{Adiabatic potential energy curves for the breathing mode of a Rydberg-eight neutral atom molecule plotted as a function of the internuclear spacing $R$. In panel (a) the neutral 
atoms (red spheres in the inset) are restricted to a plane and placed in an octagon; in (b) they are placed in a cube. In (c) the $O_h$ symmetry is broken by perturbing the atomic positions randomly by a ten percent deviation, and in (d) it is completely broken by uniformly randomly distributing the ground state atoms in a spherical shell around the Rydberg atom. The $p$-wave interaction is neglected for simplicity in (c) and (d). The colored points in (a) and (b) are the results of equation \eqref{saospwave} and in (c) and (d) of equation \eqref{highlNmanyspwave}, while the black curves are the result of diagonalization. Disagreements between the analytic and full diagonalization methods are apparent for energies between the $f$ state, with its small but non-zero quantum defect, and the hydrogenic manifold; however, for larger detunings the agreement is excellent.}
\label{trilobiteplots}
\end{center}

\end{figure*}
\noindent
The exemplary three-dimensional molecule here is a body-centered cubic, which has the highly
symmetric point group $O_h$, which decomposes into eight total irreps: 
\be
\label{Ohtrilo}
\Gamma_{O_{h}}^\text{trilobite}= A_{1g} \oplus A_{2u} \oplus F_{2u} \oplus F_{1g}
\ee
for the trilobite and $R$-butterfly, and
\be
\label{Ohbutterflies}
\Gamma_{O_h}^\text{butterflies} = E_1 \oplus E_2 \oplus F_{1g}\oplus F_{1u} \oplus F_{2g}\oplus F_{2u}
\ee
for the angular butterflies.
Figure \eqref{3dcuts} displays a series of wavefunction images highlighting the three-dimensional structure of these states. The full set of polyatomic APECs between the $n = 29$ and $n = 30$ manifolds is displayed in figure \eqref{cubepwavefull}. Compared to the diatomic case, the oscillations in the butterfly states are greatly enhanced, especially near crossings with the low-$l$ states where large wells form, strengthening the binding energies of molecules formed in these wells. 

When the neutral atoms are displaced slightly the degeneracy imposed by the $O_{h}$ group is 
broken, as shown 
in figure (\ref{trilobiteplots}(c)) and similarly in  
figure (\ref{trilobiteplots}(d)) for randomly placed atoms on a sphere.  Only the trilobite state is shown for clarity. In addition to the destruction of the degeneracy and the appearance of avoided crossings, the huge splitting between orbitals of different symmetry seen at $500\le R \le 1000$ is  reduced.

\subsection{Low-$l$ states}
\label{lowl}

\begin{figure*}[t]
\begin{center}
\subfigure[]{\label{sstate}\includegraphics[scale = 0.27]{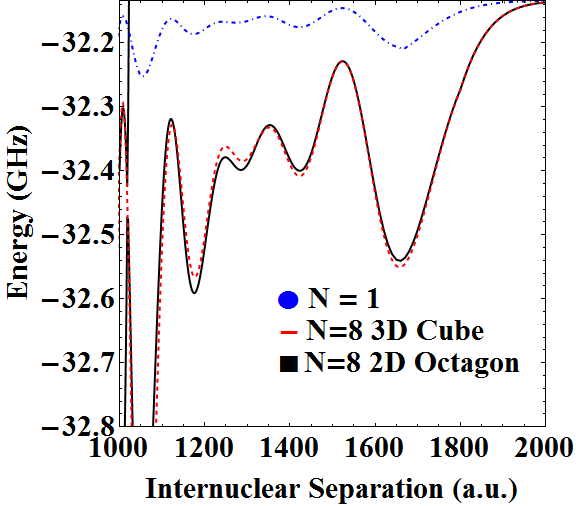}}
\subfigure[]{\label{pstate}\includegraphics[scale = 0.28]{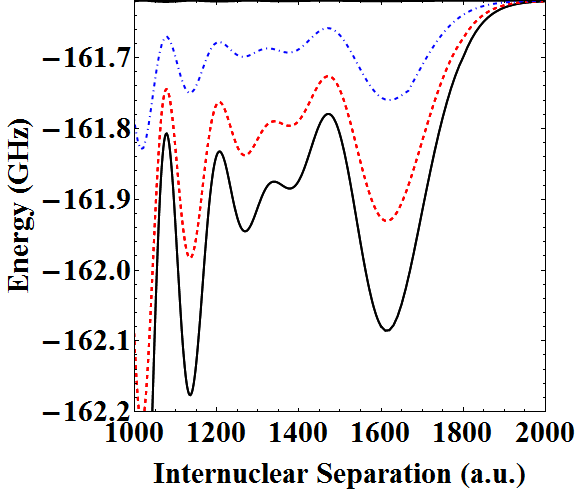}}
\subfigure[]{\label{dstate}\includegraphics[scale = 0.27]{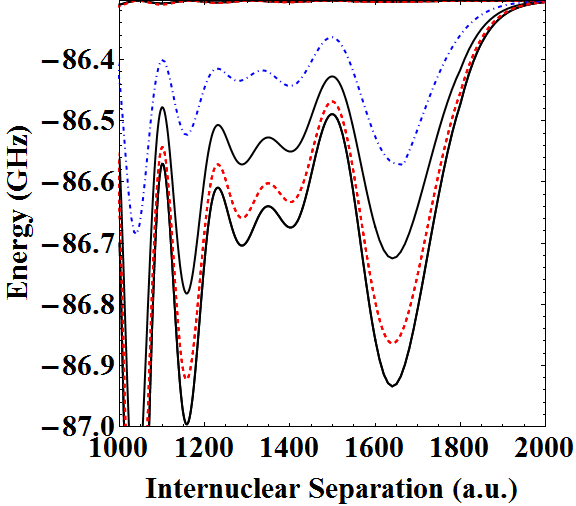}}\\
\caption{(a) $l = 0$, (b) $l = 1$, and (c) $l = 2$ low angular momentum potential energy curves for $N = 1$ (blue curve) and $N = 8$, arranged in both a coplanar octagonal geometry (black curves) and a cubic geometry (red curves). }
\label{lowlstates}
\end{center}
\end{figure*}

Most experimental probes of these exotic molecules thus far have focused on low-$l$ states. It is increasingly evident \cite{Karpiuk,GajNatComm,Schlag} that trimers, tetramers, and even pentamers are routinely formed in these experiments, and the results studied here may be relevant in explaining the non-Lorentzian line-shapes of these spectra. Although a full application of these methods to the line-shape would require investigation of the full potential energy surfaces beyond  the breathing mode cuts, some conclusions can be made. The $l = 0$ results shown in figure (\ref{lowlstates}a) are nearly independent of geometry due to the isotropy of the unperturbed state, and their depths scale linearly with $N$. According to the first order theory in equation \eqref{lowlNmany} the well depth for an $N$-atomic molecule is exactly $N$ times the diatomic depth, but due to the couplings with higher-$l$ states the depth of the largest well scales as $\simeq 0.65 N$ times the diatomic well depth for the $n = 30$ cases studied here. This scaling holds for arbitrary number of atoms and geometries.   As such, the appearance of spectral lines at integer multiples of fundamental diatomic lines signifies the production of polyatomic molecules \cite{GajNatComm}. In contrast, the $l = 1$ and $l = 2$ states are more complicated as they depend strongly on $N$ and the molecule's geometry. The spectral signatures of these polyatomic molecules will not be present as individual states, but will instead contribute to line broadening of the diatomic spectrum, and experiments with these states at high densities will need to carefully consider these effects to accurately identify spectral features. The crossover into a density shift \cite{GajNatComm,Schmidt2015} will be particularly relevant for these states.

\section{The role of density}
\label{density}

\begin{figure}[tpb]
\begin{center}
\includegraphics[scale = 0.5]{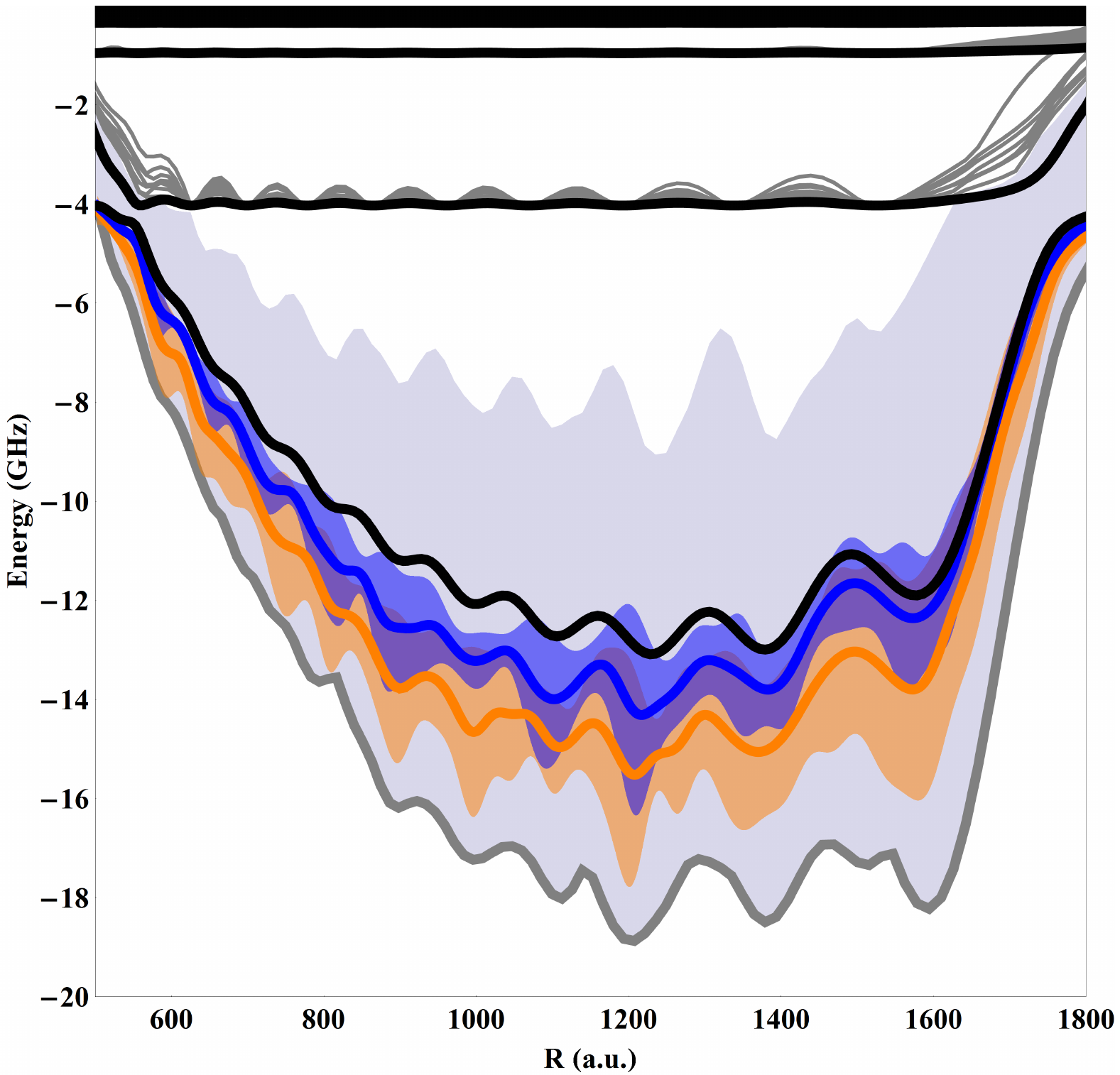}
\end{center}
\caption{ Breathing mode potential curves for the diatomic case (black), the mean (blue curve) and standard deviation (blue envelope) of 100 randomly distributed $N = 3$ systems, the mean (orange curve) and standard deviation (orange envelope) of 100 randomly distributed $N=8$ systems, and in gray the potential curves for a $N=30$ system. The overlap between orange and blue shades is denoted by a purple color. For clarity, the thirty different potential curves resulting from this case are represented with the gray envelope extending from the ground state curve to the most shallow.  }
\label{manyperturbers}
\end{figure}

When the neutral ground state atoms are configured randomly general trends with increasing $N$ can be identified. The ground state APEC for polyatomic molecules with $N = 3$, $8$, and $30$ atoms placed on the surface of a sphere of radius $R$ are studied by averaging one hundred different random geometries. The mean value and the standard deviation are presented in figure \eqref{manyperturbers}. The well 
 depths increase with $N$ and the position of the minimum is shifted towards smaller distances, compensating for the fact that the electron has to interact with an increasing number of perturbers by drawing them closer where the electron has a semiclassicaly higher kinetic energy. 
 The spreading of the well depths generally increases with $N$. The gray envelope and curve
 in this figure represent all the APECs for a single thirty-atom case, rather than the mean and standard deviation of one hundred simulations. The overall spread between the ground 
state and the highest APEC are represented by a single shade for visual clarity.

The results presented in figure \eqref{manyperturbers} are the generalization of the effects seen in the low-$l$ states to trilobite and butterfly states at high density.  Both of these results will impact experiments where the shift of the Rydberg excitation 
energy is employed to estimate the density where the excitation is created \cite{GajNatComm}. 
Typically a mean field approach is assumed and the shift of the Rydberg 
state is related with the average shift of the mean field energy due to the 
presence of the Rydberg atom in the many-body system. In this picture the shift is independent of 
the Rydberg state and depends only on the density \cite{PfauBEC,Cote}. However, the findings presented here suggest 
that the presence of multiple neutrals within the Rydberg orbit will tend to deepen 
the effective Rydberg-neutral interaction, thus imposing some constraints on the validity of the mean field approach.

\section{Conclusions}

Calculations elucidating the role of symmetry and geometry in the formation of polyatomic Rydberg molecules 
at high densities have been presented.  and provide a robust framework for studying hybridized trilobite-like molecules.  The methodology developed in the present work applies to any geometrical configuration and to high Rydberg states. These represent a significant advance towards understanding spectroscopic
results in current experiments that the spectroscopic signatures of polyatomic formation will be challenging to interpret as the results are strongly determined by the molecular geometry and the presence of any symmetries in the atomic orientation. $s$-wave states are nearly independent of the system's geometry and scale linearly with $N$, but higher angular momenta depend non-trivially on the geometry and number of atoms. These dependencies have profound impacts on the density shifts and line broadening interpretations.

 The most significant limitation of the current work is that only the breathing modes are studied where the ground state atoms are all equidistant from the Rydberg core. Although the general principles gleaned from this study will likely apply to other vibrational modes, this is still a highly simplified scenario when contrasted to an ultracold gas where atoms are randomly arrayed at different distances from the Rydberg core.
 However, this seemingly unrealistic scenario may be achieved by merging the current 
 technology in optical lattices with Rydberg spectroscopy techniques. In particular, 
 tilted optical lattices~\cite{Ma-2011} can be used to generate triply occupied Mott-Insulator states; the 
 usual techniques developed in Rydberg spectroscopy will then lead to the controlled formation of Rydberg trimers, although the position of the atoms in each lattice site will still be random. This randomness can 
 be overcome by employing a rotational optical lattice~\cite{Williams-2010}, where the centrifugal force can be used to tailor a 
 more controlled geometry. Indeed, by comparing these results with those from trimers formed in a non-rotating lattice, this method can be applied to further study the influence of the geometry. Another possibility is to use current 
 hexagonal and triangular optical lattice technology~\cite{Luhmann-2014} with lattice spacing on the order of 400 nm in order to have superior control over the geometry of the Rydberg molecules. To form molecules with this internuclear spacing would require
 higher Rydberg states on the order of $n = 70$, and thus might compromise the spectroscopy of the molecular state; optical lattices with smaller lattice spacings are therefore desirable. Finally, the possibility of 
 optical micro traps~\cite{Henderson-2009,Zimmerman-2011} has to be taken into account, since these provide opportunities to 
 design very specific arrays of single-atom traps. These traps could be designed to avoid some 
 of the problems caused by the line broadening, and also to emulate the same molecule under very different geometrical 
 considerations.

\ack {This work is supported in part by the National Science Foundation under Grants PHY-1306905 and PHY-1404419.}
\appendix
\section{Symmetry-adapted orbitals}
\label{appendo}
\subsection{Example}
We first present an example of how to calculate one matrix representation of a symmetry operation. The example molecule will have four atoms arranged in a tetrahedron, belonging to the point group $T_d$. The atoms are placed equidistant from the Rydberg core at the origin and, in order from $A$ to $D$, at $(\theta,\varphi)$ values $(0,0),(b,0),(b,2\pi/3),(b,4\pi/3)$, where $b = \arccos(-1/3)$. According to equations (\ref{chibisovthetabutterfly}-\ref{chibisovphibutterfly}) the $\theta,\varphi$ butterfly orbitals are thus parallel to the unit vectors $\sin\theta\hat z-\cos\theta\cos\varphi\hat x - \cos\theta\sin\varphi\hat y$ and $\sin\varphi\hat x - \cos\varphi\hat y$, respectively. Plugging in the actual values for these angles gives the four unit vectors for both orbitals:
\begin{align*}
&\theta: -\hat x, \frac{\hat x}{3} + \frac{2\sqrt{2}}{3}\hat z,-\frac{\hat x}{6} + \frac{\hat y}{2\sqrt{3}} + \frac{2\sqrt{2}}{3}\hat z, -\frac{\hat x}{6} -\frac{\hat y}{2\sqrt{3}} + \frac{2\sqrt{2}}{3}\hat z.\\
&\varphi: -\hat y,-\hat y,\frac{\sqrt{3}}{2}\hat x + \frac{\hat y}{2},-\frac{\sqrt{3}}{2}\hat x + \frac{\hat y}{2}.
\end{align*}
As the example symmetry operation we choose one of the $C_3$ operations corresponding to a rotation about the $z$ axis by $\frac{2\pi}{3}$ radians. This cyclically rotates the three atomic labels not along the $z$ axis, so that the symmetry operation for the trilobite orbital is
\begin{align*}
\begin{pmatrix}\psi_A\\\psi_C\\\psi_D\\\psi_B\end{pmatrix} = \begin{pmatrix}1 & 0 & 0 & 0 \\ 0 & 0 & 1 & 0 \\ 0 & 0 & 0 & 1\\ 0 & 1 & 0 & 0\end{pmatrix}\begin{pmatrix}\psi_A\\\psi_B\\\psi_C\\\psi_D\end{pmatrix}.
\end{align*}
The rotation matrix in coordinate space corresponding to this symmetry operation is
\be
\begin{pmatrix}-\frac{1}{2} & -\frac{\sqrt{3}}{2} & 0 \\ \frac{\sqrt{3}}{2} & -\frac{1}{2} & 0 \\ 0 & 0 & 1\end{pmatrix}
\ee
which then must act on the orbitals. For example the $\theta$ orbital originally at $A$, $-\hat x$, is rotated to become:
\be
\begin{pmatrix}\frac{1}{2}\\\frac{\sqrt{3}}{2} \\ 0\end{pmatrix}=\begin{pmatrix}-\frac{1}{2} & -\frac{\sqrt{3}}{2} & 0 \\ \frac{\sqrt{3}}{2} & -\frac{1}{2} & 0 \\ 0 & 0 & 1\end{pmatrix}\begin{pmatrix} -1 \\ 0 \\ 0\end{pmatrix}.
\ee
The linear combination of $\theta$ and $\varphi$ butterfly orbitals at $A$ that equals $(\frac{1}{2},-\frac{\sqrt{3}}{2},0)^T$ are then solved, giving
\be
(\psi^\theta_A)'=-\frac{1}{2}\psi^\theta_A + \frac{\sqrt{3}}{2}\psi^\varphi_B.
\ee
Likewise, the rotation matrix acting on the orbital $\psi^\theta_B$ rotates it to:
\be
\begin{pmatrix}-\frac{1}{6}\\\frac{1}{2\sqrt{3}} \\ \frac{2\sqrt{2}}{3}\end{pmatrix}=\begin{pmatrix}-\frac{1}{2} & -\frac{\sqrt{3}}{2} & 0 \\ \frac{\sqrt{3}}{2} & -\frac{1}{2} & 0 \\ 0 & 0 & 1\end{pmatrix}\begin{pmatrix}\frac{1}{3} \\ 0 \\ \frac{2\sqrt{2}}{3}\end{pmatrix}.
\ee
The resultant vector is identified as the $\theta$ orbital at $C$, so the permutation of labels was sufficient here and there is no mixing of angular butterfly orbitals. In the end, the enlarged symmetry operation matrix is
\be
\begin{pmatrix}
-\frac{1}{2} & 0 & 0 & 0 & \frac{\sqrt{3}}{2} & 0 & 0 & 0 \\
0 & 0 & 1 & 0 & 0 & 0 & 0 & 0 \\
0 & 0 & 0 & 1 & 0 & 0 & 0 & 0\\
0 & 1 & 0 & 0 & 0 & 0 & 0 & 0\\
-\frac{\sqrt{3}}{2} & 0 & 0 & 0 & -\frac{1}{2} & 0 & 0 & 0\\
0 & 0 & 0 & 0& 0 & 0 & 1 & 0\\
0 & 0 & 0 & 0& 0& 0& 0& 1\\
0 & 0 & 0 & 0 & 0& 1 & 0 & 0
\end{pmatrix}.
\ee
The presence of negative or fractional values on the diagonals is what contributes to the trace of $\hat P^j$, giving different decompositions. 
\subsection{SAO coefficients}
The coefficients $\mathcal{A}_p^{(j,\alpha)}$ are given here for the example symmetries.  The rows of each matrix correspond to a given irrep $j$ labeled in the first column; each column thereafter corresponds to a diatomic orbital at $\vec R_p$. For the octagon the labeling simply proceeds around the octagon; the first four columns of the cube correspond to the upper layer ordered counter-clockwise viewed from above; the final four correspond to the bottom layer ordered identically. 

\subsection{Octagon}
\begin{align*}
\+{\mathcal{A}^{1,2,3}_{C_{8v}}}&=\left(\begin{array}{ccccccccc} 
A_1 &1 & 1 & 1& 1 & 1 & 1&1 &1\\
B_2 & 1 & -1 & 1 & -1 & 1 & -1 & 1 & -1 \\
E_1 &1 & a & 1 & 0 & -1 & -a & -1 & 0\\
E_1 &-1 & 0 & 1 & a & 1 & 0 & -1 & -a\\
E_2 & 1 & 0 & -1 & 0 & 1 & 0 & -1 & 0 \\
E_2 & 0 & 1 & 0 & -1 & 0 & 1 & 0 & -1 \\
E_3 & -1 & a & -1 & 0 & 1 & -a & 1 & 0\\
E_3 &- 1 & 0 & 1 & -a & 1 & 0 & -1 & a\end{array}\right)
\\
\+{\mathcal{A}^{4}_{C_{8v}}}&= \left(\begin{array}{ccccccccc} 
A_2 &1 & 1 & 1& 1 & 1 & 1&1 &1\\
B_1 & 1 & -1 & 1 & -1 & 1 & -1 & 1 & -1 \\
E_1 & -1 & 0 & 1 & a & 1 & 0 & -1 & -a\\
E_1 &1 & a & 1 & 0 & -1 & -a & -1 & 0\\
E_2 & 0 & 1 & 0 & -1 & 0 & 1 & 0 & -1 \\
E_2 & 1 & 0 & -1 & 0 & 1 & 0 & -1 & 0 \\
E_3 &- 1 & 0 & 1 & -a & 1 & 0 & -1 & a\\
E_3 & -1 & a & -1 & 0 & 1 & -a & 1 & 0
 \end{array}\right),
\end{align*}
$a = \sqrt{2}$. As described in section \eqref{octagon} the $\theta$-butterfly is decoupled for all co-planar molecules.
\subsection{Cube} 
\begin{align*}
\+{A_{O_h}^{1,2}} &=  \left(\begin{array}{ccccccccc} 
A_{1g} &1 & 1 & 1& 1 & 1 & 1&1 &1\\
A_{2u} &1 & -1 & 1 & -1 & -1 & 1 & -1 & 1\\
F_{1u} &1 & -1 & 1 & -1 & 1 & -1 & 1 & -1\\
F_{1u} &-1 & 0 & 1 & 0 & 1 & 0 & -1 & 0\\
F_{1u} &0 & 1 & 0 & -1 & 0 & -1 & 0 & 1\\
F_{2g} &1 & 1 & 1 & 1 & -1 & -1 & -1 & -1\\
F_{2g} &0 & -1 & 0 & 1 & 0 & -1 & 0 & 1\\
F_{2g} &1 & 0 & -1 & 0 & 1 & 0 & -1 & 0
\end{array}\right)
\end{align*}
\begin{align*}
\+{A_{O_h}^{3}} &=  \left(\begin{array}{ccccccccc} 
E_{1} &1 & 1 & 1& 1 & -1 & -1&-1 &-1\\
E_{1} &0 & 0 & 0 & 0 & 0 & 0 & 0 & 0\\
E_{2} &1 & -1 & 1 & -1 & 1 & -1 & 1 & -1\\
E_{2} &0 & 0 & 0 & 0 & 0 & 0 & 0 & 0\\
F_{1g} &b & 0 & -b & 0 & b & 0 & -b & 0\\
F_{1g} &0 & b & 0 & -b & 0 & b & -b & 0\\
F_{1g} &0 & 0 & 0 & 0 & 0 & 0 & 0 & 0\\
F_{1u} &1 & 1 & 1 & 1 & 1 & 1 & 1 & 1\\
F_{1u} &0 & -1 & 0 & 1 & 0 & 1 & 0 & -1\\
F_{1u} &1 & 0 & -1 & 0 & -1 & 0 & 1 & 0\\
F_{2g} &1 & -1 & 1 & -1 & -1 & 1 & -1 & 1\\
F_{2g} &-1 & 0 & 1 & 0 & -1 & 0 & 1 & 0\\
F_{2g} &0 & 1 & 0 & -1 & 0 & 1 & 0 & -1\\
F_{2u} &b & 0 & -b & 0 & -b & 0 & b & 0\\
F_{2u} &0 & b & 0 & -b & 0 & -b & 0 & b\\
F_{2u} &0 & 0 & 0 & 0 & 0 & 0 & 0 & 0
\end{array}\right)
\end{align*}
\begin{align*}
\+{A_{O_h}^{4}} &=  \left(\begin{array}{ccccccccc} 
E_{1} &0 & 0 & 0 & 0 & 0 & 0 & 0 & 0\\
E_{1}&1 & 1 & 1& 1 & -1 & -1&-1 &-1\\
E_{2}  &0 & 0 & 0 & 0 & 0 & 0 & 0 & 0\\
E_{2}&1 & -1 & 1 & -1 & 1 & -1 & 1 & -1\\
F_{1g} &0 & 1 & 0 &-1 & 0& -1 & 0 & 1\\
F_{1g} &1 & 0 & -1 & 0 & 1 & 0 & -1 & 0\\
F_{1g} &1 & 0 & -1 & 0 & 1 & 0 & -1 & 0\\
F_{1u} &0 & 0 & 0 & 0 & 0 & 0 & 0 & 0 \\
F_{1u} &b & 0 & -b & 0 & b & 0 & -b & 0\\
F_{1u} &0 & b & 0 & -b & 0 & b & 0 & -b\\
F_{2g} &0 & 0 & 0 & 0 & 0 & 0 & 0 & 0\\
F_{2g} &0 & b & 0 & -b & 0 & -b & 0 & b\\
F_{2g} &b & 0 & -b & 0 & -b & 0 & b & 0\\
F_{2u} &0 & -1 & 0 & 1 & 0 & -1 & 0 & 1\\
F_{2u} &1 & 0 & -1 & -0 & 1 & 0 & -1 & 0\\
F_{2u} &1 & -1 & 1 & -1 & -1 & 1 & -1 & 1
\end{array}\right),
\end{align*}
$b = \sqrt{3}$. 
 \section*{References}
 

\end{document}